\documentclass[trackchanges,twocolumn]{aastex701}
\usepackage{CJKutf8}
\usepackage{amsmath}

\newcommand{\namecn}[1]{\begin{CJK*}{UTF8}{gbsn}({#1})\end{CJK*}}

\begin{document}

\title{Detection of dark companions via the combination of eclipse timing variation, Hipparcos and/or Gaia astrometry: the cases of V Puppis and CY Ari}

\author[orcid=0000-0001-6753-4611]{Guang-Yao Xiao \namecn{肖光耀}}
\affiliation{State Key Laboratory of Dark Matter Physics, Tsung-Dao Lee Institute \& School of Physics and Astronomy, Shanghai Jiao Tong University, Shanghai 201210, China}
\email[show]{gyxiao{\_}tdli@sjtu.edu.cn}  


\author[orcid=0000-0001-6039-0555]{Fabo Feng \namecn{冯发波}}
\affiliation{State Key Laboratory of Dark Matter Physics, Tsung-Dao Lee Institute \& School of Physics and Astronomy, Shanghai Jiao Tong University, Shanghai 201210, China}
\email[show]{ffeng@sjtu.edu.cn}

\author[orcid=0000-0003-3116-5038]{Song Wang \namecn{王松}}
\affiliation{Key Laboratory of Optical Astronomy, National Astronomical Observatories, Chinese Academy of Sciences, Beijing, China}
\affiliation{Institute for Frontiers in Astronomy and Astrophysics, Beijing Normal University, Beijing, 102206, China}
\email{songw@bao.ac.cn}

\author[orcid=0000-0003-3590-335X]{Kai Li \namecn{李凯}}
\affiliation{Shandong Key Laboratory of Space Environment and Exploration Technology, Institute of Space Sciences, School of Space Science and Technology, Shandong University, Shandong, China}
\email{kaili@sdu.edu.cn}


\author[orcid=0000-0001-5295-1682]{Yicheng Rui \namecn{芮易成}}
\affiliation{State Key Laboratory of Dark Matter Physics, Tsung-Dao Lee Institute \& School of Physics and Astronomy, Shanghai Jiao Tong University, Shanghai 201210, China}
\email{ruiyicheng@sjtu.edu.cn}

\author[orcid=0000-0002-6573-6719]{Xiao-Wei Duan \namecn{段晓苇}}
\affiliation{State Key Laboratory of Dark Matter Physics, Tsung-Dao Lee Institute \& School of Physics and Astronomy, Shanghai Jiao Tong University, Shanghai 201210, China}
\email{ruiyicheng@sjtu.edu.cn}


\begin{abstract}

The third body is expected to shape the formation and evolution of close binary systems. In this work, we develop a method to detect and characterize the tertiary companion around eclipsing binaries through the combined analysis of eclipse timing variation, Hipparcos and/or Gaia astrometry. This method allows us to determine both the true mass and the inclination of the tertiary body that inferred from light-travel time effect.
For the massive B-type binary V Pup, we do not confirm the previously reported 5.47-yr signal; instead, we identify a longer period of 14 yr. The orbital semi-major axis and mass of the outer body are revised to $a_C={17.88}_{-0.15}^{+0.15}$\,au and $M_C={7.73}_{-0.14}^{+0.14}\,M_\odot$, confirming it as a promising stellar-mass black-hole candidate for further follow-up study.  
For the tertiary of the contact binary CY Ari, we obtain $P_C=5.406_{-0.016}^{+0.017}$ yr, $e_C=0.526_{-0.027}^{+0.032}$, $I_C={85.6}_{-6.5}^{+7.8}$$^\circ$, and a true mass of $M_C=0.640_{-0.029}^{+0.029}\,M_\odot$, supporting the white dwarf hypothesis proposed in previous study. 
The orbits of both systems are nearly edge-on ($I=90^{\circ}$), implying that they may form in a coplanar environment. We highlight the advantages of our method for detecting dark companions in binary and triple systems.

\end{abstract}



\section{Introduction}

The eclipse time variation (ETV) technique, also known as O-C (observed minus calculated) analysis, has been extensively applied to investigate period variations in eclipsing binary (EB) systems. It also helps to reveal the properties and evolutionary states of EBs that host a tertiary body. 

ETVs can originate from a single mechanism or from a combination of multiple causes; they may be physical, reflecting changes in the real orbital period, or merely apparent and lacking any underlying physical origin. 
The well-understood origins of ETVs include long- and short-term physical variations, as well as erratic variations. 
Long-term variations generally arise from mass transfer between binary components, mass loss driven by stellar wind, magnetic braking \citep{Verbunt1981A&A}, and tidal dissipation. Because the characteristic timescales of these processes exceed typical observational baselines of ETVs, the O-C diagram appears as a clear parabola opening either upward or downward (e.g., \citealt{Nanouris2015A&A}). 
The short-term ETVs typically manifest as periodic oscillations, involving magnetic activity (e.g., Applegate mechanism, \citealt{Applegate1992ApJ}), the light-travel time effect (LTTE), short-period dynamical perturbations, or the apsidal motion effect (AME). 
The LTTE is caused by the changing distance of a binary in a hierarchical multiple-star system and usually dominates the ETVs for close binaries with distant companions. For eccentric binaries or more compact triples, however, AME and dynamical interaction between the inner pair and the outer body become non-negligible (e.g., \citealt{Borkovits2015MNRAS, Borkovits2016MNRAS, Mitnyan2024A&A, Borkovits2025A&A}). 

In this work, we focus on the LTTE analysis. As with the radial velocity (RV) technique, the LTTE-only solution also suffers from the so-called $M_{\rm }\,{\rm sin}\,I$ degeneracy, where $I$ is the orbital inclination of the third body. It means that LTTE method can only measure a minimum mass instead of true mass of the companion. This degeneracy can be addressed by including the dynamical terms in the ETV model (e.g.,\citealt{Borkovits2015MNRAS}) or by incorporating additional constraints from astrometry. 


In the Kepler\citep{Borucki2010Sci} and TESS \citep{Ricker2015} era, precise space photometry has uncovered thousands of EBs and hundreds of hierarchical triples. Numerous groups have subsequently carried out comprehensive studies of the mechanisms that drive ETVs (e.g., \citealt{Rappaport2013ApJ,Conroy2014AJ,Gies2015AJ,Borkovits2016MNRAS,Inacio2024MNRAS}). Astrometric measurements from Gaia or Hipparcos can also provide independent or complementary evidence to detect or confirm the tertiary body. For example, \citet{Mitnyan2024A&A} validated their ETV solutions against Gaia non-single star (NSS) orbital solutions and found consistent parameters for several dozen hierarchical-triple candidates. However, for long-period candidates, NSS results are still limited by Gaia's $\sim1000$-day baseline.

\begin{table*}[!]
\centering
\caption{Properties of host stars for V Pup, CY\,Ari and Gaia BH3 system}\label{Tab:stellar}
\begin{tabular*}{\textwidth}{@{}@{\extracolsep{\fill}}lcccc@{}}
\hline \hline
Parameter & V Pup &{CY Ari} & {Gaia BH3} & Reference \\
\hline
\textbf{Gaia DR3}&\\
ID&5517171678276669696&{101358960442870528}&{ 4318465066420528000}&1\\
$\alpha$ (deg)$\rm ^a$&119.5601247625 (0.16)&36.5057954842 (0.028)&294.8278625082 (0.051)&1\\
$\delta$ (deg)$\rm ^a$&$-$49.24487132664 (0.16)&22.4458266130 (0.034)&14.9309796086 (0.052)&1\\
$\varpi$ (mas)&$2.72\pm 0.19$&$2.966\pm0.030$&$1.679\pm0.069$&1\\
$\mu_{\alpha\ast}$ ($\rm mas\,yr^{-1}$)&$-5.28\pm 0.26$&$27.299\pm0.036$&$-22.235\pm0.062$&1\\
$\mu_{\delta}$ ($\rm mas\,yr^{-1}$)&$9.87 \pm0.25$&$-13.134\pm0.039$&$-155.276\pm0.059$&1\\
\hline
\textbf{Gaia DR2}&\\
$\alpha$ (deg)$\rm ^a$&119.5601262544 (0.58)&36.5057913807 (0.041)&294.8278656955 (0.027)   &2\\
$\delta$ (deg)$\rm ^a$&$-$49.2448726030 (0.72)&22.4458284393 (0.031)&14.9310011330 (0.025)  &2\\
$\varpi$ (mas)&$0.89 \pm0.73$&$2.951\pm0.042$&$1.630\pm0.037$&2\\
$\mu_{\alpha\ast}$ ($\rm mas\,yr^{-1}$)&$-5.9 \pm1.1$&$26.708\pm0.088$&$-22.214\pm0.053$&2\\
$\mu_{\delta}$ ($\rm mas\,yr^{-1}$)&$12.6\pm 1.2$&$-12.777\pm0.073$&$-156.765\pm0.053$&2\\
\hline
$G$ (mag)&4.4602&{11.7601}&{11.2310}&1\\
RUWE&2.41&1.88&3.41&1\\
$T_{\rm eff}\,(\rm K)$ &$26000\pm1000$ ($T_A$); $24000\pm1000$ ($T_B$)&{$5851$}&{$5212\pm80$}&3, 4, 5\\
${\rm log}\,\textsl{g}$ (cgs) &---&{$4.298$}&{$2.929\pm0.003$}&3, 4, 5\\
$\rm [Fe/H]$ (dex) &---&{$0.173$}&{$-2.56\pm0.11$}&3, 4, 5\\
$M_{\star}\,(M_{\sun})$$\rm ^b$ &$14.0\pm0.5$ ($M_A$); $7.3\pm0.3$ ($M_B$) &{$1.1\,(M_A)$; $0.3\,(M_B)$}&{$0.76\pm0.05$}&3, 4, 5\\
$R_\star\,(R_{\sun})$&$5.48\pm0.18$ ($R_A$); $4.59\pm0.15$ ($R_B$)&{$1.3\,(R_A)$; $0.7\,(R_B)$}& $4.936\pm0.016$&3, 4, 5\\
log$(L_\star/L_{\sun})$&$4.10\pm0.08$ ($L_A$); $3.81 \pm0.08$ ($L_B$) &{$0.20\,(L_A)$; $-0.22\,(L_B)$}&$1.208\pm0.030$&3, 4, 5\\
\hline
\multicolumn{5}{l}{$\rm ^a$ The figures between brackets denote the uncertainties of $\alpha$ and $\delta$ in mas unit.}\\
\multicolumn{5}{l}{$\rm ^b$ The subscript ``A'' labels the the more massive component, while ``B'' denotes the less massive one in the inner binary.}\\
\multicolumn{5}{l}{\textbf{References}$-$(1)\citet{GaiaCollaboration2023}; (2)\citet{GaiaCollaboration2018}; (3)\citet{Xu_CYAri_2025ApJ};}\\
\multicolumn{5}{l}{(4)\citet{GaiaBH3_2024}; (5)\citet{Budding2021MNRAS}.}
\end{tabular*}
\end{table*}

In this work, we present a joint analysis of TTV (mainly LTTE), Hipparcos intermediate astrometry data (IAD; \citealt{vanLeeuwen2007}), and/or Gaia second and third data releases (GDR2 and GDR3; \citealt{GaiaCollaboration2018,GaiaCollaboration2023}) to constrain the orbital parameters and mass of the third component orbiting the EBs. For targets with Hipparcos and Gaia measurements, the long-term proper motion anomaly (PMa) with a 25-year span can provide additional constraint for a long-period orbit (e.g., \citealt{Brandt2018,Kervella2019,Kervella2022,Feng2022,Xiao2023}). For systems with only Gaia astrometry, we use both GDR2 and GDR3 catalog astrometry, which can also extend the sensitive orbital period of the third body to thousands of days.
We perform a comparative analysis of the well-known Gaia BH3 system to verify the robustness of our method (mainly the astrometric part). Then we take V Pup (with Hipparcos) and CY Ari (without Hipparcos) as examples to demonstrate the fitting results from our joint analysis. 

V Pup (HIP38957) is a bright eclipsing variable ($P=1.45$ day) containing two massive B-type components. According to LTTE-only analysis, \citet{Qian2008ApJ} reported the presence of a massive black hole candidate ($M>10.4\,M_\sun$, $P=5.47$ yr) as a tertiary companion to this system.
CY Ari comprises a W UMa-type contact binary ($P=0.38$ day) and a faint, wide companion ($P=5.41$ yr) recently identified by a LTTE-only analysis \citep{Xu_CYAri_2025ApJ}. This system exhibits a significant astrometric signature within the difference between GDR2 and GDR3. The stellar and astrometric parameters of these systems are summarized in Table~\ref{Tab:stellar}.

\section{Method}
We adopt a right-handed coordinate system defined by the orthonormal triad $[\hat{x}, \hat{y}, \hat{z}]$, where $\hat{x}$ points the direction of increasing R.A., $\hat{y}$ points the direction of increasing decl., and $\hat{z}$ points away from the observer (convention $\mathrm{III}$ of \citealt{Feng2019ApJS..242...25F}). 
This convention is used in the astrometry models of Hipparcos and Gaia \citep{ESA1997, Lindegren2012A&A}. 

\subsection{ETV model}
For a typical triple-star system that consists of a close EB and a more distant outer body, the most general form of the ETV model includes a polynomial term, the LTTE term ($\Delta_{\rm LTTE}$, or R$\o$mer delay), $P_2$ timescale dynamical perturbations ($\Delta_{\rm dyn}$), and apsidal motion ($\Delta_{\rm apse}$), which can be expressed as \citep{Borkovits2015MNRAS} 
\begin{equation}
\Delta \hat{T_E}=\sum^3_{i=0}c_iE^i+[\Delta_{\rm LTTE}+\Delta_{\rm dyn}+\Delta_{\rm apse}]_0^E,
\end{equation}
where $E$ represents the $E$th cycle of the a specific eclipse minimum relative to the reference epoch. The polynomial term contains corrections for the calculated times of eclipse minima ($i=0,1$), a steady period change (quadratic term, $i=2$), and a constant period acceleration (cubic term, $i=3$). In this work, we restrict our analysis to the quadratic term, and therefore, the above equation can be rewritten as a more specific form, i.e.,
\begin{equation}
\begin{split}
\Delta \hat{T_E}=
&\Delta T_0 + \Delta P_0 \cdot E + \frac{1}{2}\frac{dP}{dE}E^2\\
&+[\Delta_{\rm LTTE}+\Delta_{\rm dyn}+\Delta_{\rm apse}]_0^E,
\end{split}
\end{equation}
where $\Delta T_0$ is the correction for the reference time $T_0$, $\Delta P_0$ is the correction for the reference eclipsing binary period $P_0$, and $\dot{P}$ is the time derivation of the period, which is usually interpreted as mass transfer between the components, angular momentum loss, and/or thermal relaxation oscillations (e.g., \citealt{Qian2001MNRAS}). Because the ETV timing signal is typically much less than the eclipsing binary period, $E$ is approximately $(T_E-T_0)/P_0$. Hence the above equation can also be written as
\begin{equation}
\begin{split}
\Delta \hat{T_E}=
&\Delta T_0 + g\,(T_E-T_0)+h\,(T_E-T_0)^2\\
&+[\Delta_{\rm LTTE}
+\Delta_{\rm dyn}+\Delta_{\rm apse}]_0^E,
\end{split}
\end{equation}
where $T_E$ is the observed eclipse time for the $E$th cycle, $g\equiv\Delta P_0/P_0$, and $h\equiv\dot{P}/(2P_0)$.

The LTTE arises because the distance between the inner binary and the observer varies periodically as the EB orbits the common center of mass of the triple system, producing a corresponding periodic shift in light travel time. This time shift or delay can be modeled as
\begin{equation}
\label{eq:LTTE}
\Delta_{\rm LTTE} = - \frac{a_{\rm AB}\,{\rm sin}I_2}{C}\frac{(1-e_2)^2\,{\rm sin}(\nu_2+\omega_2)}{1+e_2\,{\rm cos}\,\nu_2},
\end{equation}
where $a_{\rm AB}$ is the semi-major axis of the inner binary's barycenter relative to the barycenter of the whole system (BOS for short), $C$ is the speed of light, $I_2$, $e_2$, $\nu_2$ and $\omega_2$ are the inclination, eccentricity, true anomaly and the argument of periastron of the orbit of tertiary body. If we use the orbital elements of the reflex motion of the eclipsing binary induced by the outer companion for the last four quantities, the sign ``-'' in front of the equation should be revised to ``+''. 

In addition to LTTE, $P_2$ timescale dynamical perturbations stem from gravitational coupling between the outer companion and the inner binary, which can alter the orbital elements of the binary and thereby induce measurable period variations \citet{Borkovits2011A&A}. However, for a close binary with a more wider companion, this effect will be greatly weakened. Following \citet{Mitnyan2024A&A} and \citet{Borkovits2025A&A}, we calculate the amplitude ratios $\mathcal{A}_{\rm dyn}/\mathcal{A}_{\rm LTTE}$ and $\mathcal{A}_{\rm dyn}^{\rm coplanar}/\mathcal{A}_{\rm LTTE}$ to quantify its contribution. The amplitude of LTTE is 
\begin{equation}
\mathcal{A}_{\rm LTTE} = \frac{a_{\rm AB}\,{\rm sin}I_2}{C}\sqrt{1-e_2^2\,{\rm cos^2}\omega_2}\,,
\end{equation}
and the characteristic amplitude of the dynamical effect is 
\begin{equation}
\mathcal{A}_{\rm dyn}=\frac{1}{2\pi}\frac{m_C}{m_{ABC}}\frac{P_1^2}{P_2}(1-e_2^2)^{-3/2},
\end{equation}
where $m_C$ and $m_{ABC}$ are the mass of the outer companion and the total mass of the whole system, and $P_1$ and $P_2$ are the period of the binary and outer companion. respectively. For nearly coplanar orbital configuration between inner binary and outer companion, the above equation is modified as \citep{Rappaport2013ApJ}
\begin{equation}
\mathcal{A}_{\rm dyn}^{\rm coplanar}=\frac{3}{2\pi}\frac{m_C}{m_{ABC}}\frac{P_1^2}{P_2}(1-e_2^2)^{-3/2}\,e_2.
\end{equation}
If the ratios are higher than $\sim0.3$, the dynamical effect can not be ignored. 
For the V Pup system, these two ratios are determined to be $\mathcal{A}_{\rm dyn}/\mathcal{A}_{\rm LTTE}=0.0010$ and $\mathcal{A}_{\rm dyn}^{\rm coplanar}/\mathcal{A}_{\rm LTTE}=0.0013$, and for the CY Ari system, they are 0.0008 and 0.0013, respectively, suggesting that the dynamical effect is negligible.

Apsidal motion commonly occurs in eccentric eclipsing binaries and is driven by general-relativistic precession, classical tidal distortions and any additional dynamical perturbations. Following the formulation of \citet{Mitnyan2024A&A} (Equation 15), the apsidal-motion term vanishes for a close binary whose orbit has been circularized. Therefore, we only consider the contribution from LTTE for the two systems. An LTTE-only model provides the same information as a RV-only solution and suffers from the same $M\sin I$ degeneracy.

The likelihood of ETV (O-C) can be calculated by
\begin{equation}
  \mathcal{L}_{\rm ETV}=\prod\limits_{E}^{N_{\rm ETV}}\frac{1}{\sqrt{2\pi\sigma_{T_E}^2}}{\rm exp}\left(-\frac{(\Delta T_E-\Delta\hat{T_E})^2}{2\sigma_{T_E}^2}    \right)~,
\end{equation}
where $\Delta T_E\equiv T_E-(T_0+P_0E)$ and $\sigma_{T_E}$ are the calculated O-C based on the observed eclipse minimum time and associated uncertainty, respectively. To keep consistency with the RV model defined in our previous work (see the Appendix of \citealt{Xiao2024MNRAS}), we focus on stellar reflex motion (precisely, it is the motion of the inner binary's barycenter around BOS) and choose the orbital period $P$, RV semi-amplitude $K$, eccentricity $e$, argument of periastron $\omega$, mean anomaly $M_{0}$ at the reference epoch as the free parameters of LTTE-only analysis. They can be easily used  to derive Equation~\ref{eq:LTTE}. The additional three parameters are $\Delta T_0$, $\Delta P_0$ and $\dot{P}$.

\subsection{Gaia DR2 and DR3 model}
Since the Gaia astrometric epoch data is not available, we use Gaia Observation Forecast Tool \footnote{\url{https://gaia.esac.esa.int/gost/index.jsp}} (GOST) to simulate the Gaia Along-scan (AL) abscissa for GDR2 and GDR3, respectively. The abscissa will take into account the linear motion of BOS, the reflex motion of the host star induced by outer companions, and the parallax motion of the satellite. We will fit a standard five-parameter astrometric model to the abscissa and compare the fitted values to the catalog astrometry of Gaia archive. With a baseline of about half a year between GDR2 and GDR3, any significant deviation from linear motion will appear as an astrometric anomaly, implying the presence of unsolved companions.

In rectangular coordinates, the Thiele-Innes coefficients $A$, $B$, $F$, $G$ \citep{ Binnendijk1960, Heintz1978, Thiele1883} are defined as 
\begin{equation}
    A={\rm cos}\,\omega\,{\rm cos}\,\Omega -{\rm sin}\,\omega\,{\rm sin}\,\Omega\,{\rm cos}\,i,
\end{equation}
\begin{equation}
    B={\rm cos}\,\omega\,{\rm sin}\,\Omega +{\rm sin}\,\omega\,{\rm cos}\,\Omega\,{\rm cos}\,i,
\end{equation}
\begin{equation}
    F=-{\rm sin}\,\omega\,{\rm cos}\,\Omega -{\rm cos}\,\omega\,{\rm sin}\,\Omega\,{\rm cos}\,i,
\end{equation}
\begin{equation}
    G=-{\rm sin}\,\omega\,{\rm sin}\,\Omega +{\rm cos}\,\omega\,{\rm cos}\,\Omega\,{\rm cos}\,i.
\end{equation}
Besides, the elliptical rectangular coordinates $X$ and $Y$ are functions of the eccentric anomaly $E(t)$ and the eccentricity $e$, which are given by
\begin{equation}
    X={\rm cos}\,E(t)-e
\end{equation}
\begin{equation}
    Y=\sqrt{1-e^2}\cdot {\rm sin}\,E(t).
\end{equation}
Therefore, the projected offsets of the stellar reflex motion relative to the BOS are then given by
\begin{equation}
    \Delta\alpha^{r}_\ast=a_{\rm phot}\,\varpi\,(BX+GY),
\end{equation}
\begin{equation}
     \Delta\delta^{r}=a_{\rm phot}\,\varpi\,(AX+FY),
\end{equation}
where $\Delta\delta^{r}$ and $\Delta\alpha^{r}_\ast=\Delta\alpha^r\,{\rm cos}\,\delta^r$ are the offsets in declination and right ascension, respectively, and $\varpi$ is the parallax in units of mas. $a_{\rm phot}$ is the semi-major axis of the system's photocentre relative to BOS. It can be written as

\begin{equation}
\label{equ:photo}
a_{\rm phot} = a_\star(1-\frac{m_\star\,f_c}{m_c\,f_\star})(1+\frac{f_c}{f_\star})^{-1},  
\end{equation}
where $a_\star$ is the semi-major axis of the host star (or $a_{\rm AB}$ for triples) relative to BOS, $m_\star$ and $m_c$ are the masses of primary (or binary mass $m_{\rm AB}$) and companion, and $f_\star$ and $f_c$ are their fluxes in Gaia G band. If the flux ratio is unknown, we interpolate the mass-absolute G magnitude (M-G) relation of \citet{mamajek2013ApJS} \footnote{\url{https://www.pas.rochester.edu/~emamajek/EEM_dwarf_UBVIJHK_colors_Teff.dat}} to derive it. If the companion is quite fainter than the host, its luminosity contributed to the system's photocentre is negligible, i.e., $a_{\rm phot}=a_\star$.  Assuming the outer companion of the CY Ari system ($m_A\sim1.1\,M_\odot, m_B\sim0.3\,M_\odot$, and $m_C\sim~0.64\,M_\odot$) is a main-sequence star, we find $a_{\rm phot}\sim0.92\,a_\star$, which appears to have no significant impact on the final orbital solution. Noted that we have ignored the photocentre motion of the inner binary around the binary barycenter. For example, in the case of CY Ari AB, the above equation gives an angular separation of only $a^{\rm phot}_{\rm binary}\sim0.007$ mas between the photocentre and the barycenter, well below the detection limit of Gaia. Moreover, both components of CY Ari have evolved to fill their critical Roche lobes, so the separation is small enough that the photocentre offset can be neglected. However, for nearby systems with relatively long binary periods, this correction is necessary.  

Next, we model the astrometry of the BOS at the GDR3 epoch ($t_{\rm DR3}={\rm J}2016.0$) as follows \citep{Feng2023MNRAS}:
\begin{equation}
    \alpha^{b}_{\rm DR3}=\alpha_{\rm DR3}-\frac{\Delta\alpha_\ast}{{\rm cos}\,\delta_{\rm DR3}}
\end{equation}
\begin{equation}
    \delta^{b}_{\rm DR3}=\delta_{\rm DR3}-\Delta\delta
\end{equation}
\begin{equation}
    \varpi^{b}_{\rm DR3}=\varpi_{\rm DR3}-\Delta\varpi
\end{equation}
\begin{equation}
    \mu^{b}_{\alpha{\rm DR3}}=\mu_{\alpha{\rm DR3}}-\Delta\mu_{\alpha}
\end{equation}
\begin{equation}
    \mu^{b}_{\delta{\rm DR3}}=\mu_{\delta{\rm DR3}}-\Delta\mu_{\delta},
\end{equation}
where $\alpha$, $\delta$, $\mu_{\alpha}$, $\mu_{\delta}$ are right ascension, declination and corresponding proper motions, and the subscript $_{\rm DR3}$ and the superscript $^b$ represent quantities of GDR3 and BOS astrometry, respectively.
Above five quantities with $\Delta$ are barycenter offsets relative to GDR3 catalog astrometry, and will be set as free parameters.
The BOS astrometry at reference epoch $t_k$ ($k=1,2$ represent GDR2, GDR3) can be modeled through a simple linear propagation.

By combining the linear motion of BOS and the stellar reflex motion, we can simulate the Gaia AL abscissae directly. For GDR3 abscissae, we first simulate the position of the target at GOST epoch $t_j$ relative to $t_{\rm DR3}$ using 
\begin{equation}
    \Delta\alpha_{\ast{j}}=\Delta\alpha^{b}_{\ast{\rm DR3}}+\mu^{b}_{\alpha {\rm DR3}}(t_j-t_{\rm DR3})+\Delta\alpha^{r}_{\ast j}
\end{equation}
\begin{equation}
    \Delta\delta_{j}=\Delta\delta^{b}_{\rm DR3}+\mu^{b}_{\delta {\rm DR3}}(t_j-t_{\rm DR3})+\Delta\delta^{r}_{j},
\end{equation}
where $\Delta\alpha^{b}_{\ast{\rm DR3}}=(\alpha^b_{\rm DR3}-\alpha_{\rm DR3})\,\cos\delta^b_{\rm DR3}$, and $\Delta\delta^{b}_{\rm DR3}=\delta^b_{\rm DR3}-\delta_{\rm DR3}$. Since the reflex motion induced by outer companions is not as significant as linear barycentric motion, we approximate the parallax at $t_j$ as $\varpi_j\approx\varpi^b_{\rm DR3}$. Then we project the above target position onto the 1D AL direction by considering the parallactic perturbation of Gaia satellite's heliocentric motion, using 
\begin{equation}
    \eta_j = \Delta\alpha_{\ast{j}}\,{\rm sin}\,\psi_j+\Delta\delta_{j}\,{\rm cos}\,\psi_j+\varpi^{b}_{\rm DR3}f^{\rm AL}_j,~
    \label{equ:gaia_abs}
\end{equation}
where $\eta_j$ is AL abscissa, $\psi_j$ is the scan angle of Gaia satellite, and $f^{\rm AL}_j$ is the parallax factor from GOST. Finally, we model the simulated abscissae with a standard five-parameter astrometric model (or single-source model) as follows:
\begin{equation}
    \hat{\eta_j} = \Delta\alpha^{l}_{\ast{\rm DR3}}\,{\rm sin}\,\psi_j+\Delta\delta^{l}_{\rm DR3}\,{\rm cos}\,\psi_j+\hat{\varpi}_{\rm DR3}f^{\rm AL}_j,
\label{5-p_ast}
\end{equation}
\begin{equation}
\Delta\alpha^{l}_{\ast{\rm DR3}}=(\hat{\alpha}_{\rm DR3}-\alpha_{\rm DR3})\,{\rm cos}\,\hat{\delta}_{\rm DR3}+\hat{\mu}_{\alpha{\rm DR3}}(t_j-t_{\rm DR3}),
\end{equation}
\begin{equation}
\Delta\delta^{l}_{\rm DR3}=(\hat{\delta}_{\rm DR3}-\delta_{\rm DR3})+\hat{\mu}_{\delta{\rm DR3}}(t_j-t_{\rm DR3}).
\end{equation}
Above modeling can give a set of model parameters ($\hat{\alpha}_{\rm DR3}, \hat{\delta}_{\rm DR3}, \hat{\mu}_{\alpha{\rm DR3}},   \hat{\delta}_{\rm DR3}, \hat{\varpi}_{\rm DR3}$) at $t_{\rm DR3}$. Likewise, modeling GDR2 astrometry can be done easily by changing the subscript $_{\rm DR3}$ to $_{\rm DR2}$, but keeping the reference position fixed in GDR3. Given that the Gaia epoch data is not available, we assume each individual measurement has the same uncertainty and thus will be assigned equal weighting when fitting for the five-parameter model. 

To avoid numerical errors, we define the catalog astrometry at $t_k$ relative to $t_{\rm DR3}$ as follow:
\begin{equation}
\begin{split}
\Delta\vec{\iota}_k
&\equiv\,(\Delta\alpha_{\ast k},\,\Delta\delta_k,\,\Delta\varpi_k,\,\Delta\mu_{\alpha k},\,\Delta\mu_{\delta k})\\
&=((\alpha_k-\alpha_{\rm DR3})\,{\rm cos\delta_k},\,\delta_k-\delta_{\rm DR3},\,\varpi_k-\varpi_{\rm DR3},\\
&\,\,\,\,\,\,\,\,\mu_{\alpha k}-\mu_{\alpha \rm DR3},\,\mu_{\delta k}-\mu_{\delta \rm DR3}).
\end{split}
\end{equation}
Likewise, the fitted astrometry at $t_k$ is $\Delta\hat{\vec{\iota}}_k$. The likelihood for GDR2 and GDR3 can be written as
\begin{equation}
\begin{split}
  \mathcal{L}_{\rm DR}=
  &\prod\limits_{k=1}^{N_{\rm DR}}\frac{1}{\sqrt{(2\pi)^5|\Sigma_k(S^2)|}}\times\\
  &{\rm exp}\left(-\frac{1}{2}(\Delta\hat{\vec{\iota}}_k-\Delta\vec{\iota}_k)^{T}[\Sigma_k(S^2)]^{-1}(\Delta\hat{\vec{\iota}}_k-\Delta\vec{\iota}_k)    \right)~,
\end{split}
\end{equation}
where $N_{\rm DR}$ represents the number of Gaia data releases ($N_{\rm DR}=2$ if we use both GDR2 and GDR3), $\Sigma_k$ is the catalog covariance for the five parameters, and $S$ is the error inflation factor for Gaia astrometry. More details are presented in \citet{Xiao2024MNRAS}. For GDR23 model, the addition free parameters are orbital inclination $I$, longitude of ascending node $\Omega$, five astrometric offsets ($\Delta \alpha*$, $\Delta \delta$, $\Delta \varpi$, $\Delta \mu_{\alpha*}$ and $\Delta \mu_\delta$) of barycenter relative to GDR3, and the error inflation factor $S$.

\subsection{Gaia epoch astrometry model}
The Gaia epoch astrometry, derived from 66 months of observations, will be released not earlier than December 2026 in Gaia DR4. It will provide the observational barycentric time, the AL position of the photocentre, the scan angle, and the parallax factor for each transit of a source. Similar to Equation~\ref{equ:gaia_abs}, the Gaia AL abscissa of a typical two-body system can be modelled by the combined form of a single-source model, which describes the motion of BOS, and a Keplerian model that describes the stellar reflex motion \citep{Holl2023A&A}. The general form is  
\begin{equation}
\begin{split}
    \hat{\eta}_j = 
    &(\Delta\alpha_{\ast{}}^b+\mu^b_{\alpha}\cdot\Delta t)\,{\rm sin}\,\psi_j+(\Delta\delta^{b}+\mu_\delta^b\cdot\Delta t)\,{\rm cos}\,\psi_j\\
    &+\varpi^{b}_{}f^{\rm AL}_j+\Delta\alpha^{r}_{\ast j}\,{\rm sin}\,\psi_j + \Delta\delta^{r}_{j}\,{\rm cos}\,\psi_j,
\end{split}
\end{equation}
where ($\Delta\alpha_{\ast{}}^b$, $\Delta\delta^{b}$) are the relative offsets of BOS relative to a reference position, ($\mu^b_{\alpha}$, $\mu_\delta^b$, $\varpi^{b}$) are the absolute proper motions and parallax of BOS, and $\Delta t=t_j-t_{\rm DR4}$ is the time relative to DR4 reference epoch (J2017.5 or JD 2457936.875).

The likelihood is
\begin{equation}
  \mathcal{L}_{\rm DR4}=\prod\limits_{j=1}^{N_{\rm DR4}}\frac{1}{\sqrt{2\pi(\sigma_{j}^2+J_{\rm gaia}^2)}}{\rm exp}\left(-\frac{(\hat{\eta}_j-\eta_j)^2}{2(\sigma_{j}^2+J_{\rm gaia}^2)}    \right)~,
\end{equation}
where $N_{\rm DR4}$ is the total number of Gaia epoch astrometric data, $\sigma_j$ is the uncertainty of the individual measurement and $J_{\rm gaia}$ is the additional jitter term. In principle, Gaia epoch astrometry is capable of determining the full orbital elements of a Keplerian motion.

\subsection{Hipparcos epoch astrometry model}
As in the GDR23 model, we first propagate the BOS astrometry from GDR3 reference epoch $t_{\rm DR3}$ to Hipparcos reference epoch $t_{\rm HIP}$. Then we simulate the position of target at Hipparcos epoch using
\begin{equation}
    \Delta\alpha_{\ast{j}}=\Delta\alpha^b_{\ast{\rm HIP}}+\Delta\mu^b_{\alpha {\rm HIP}}(t_j-t_{\rm HIP})+\Delta\alpha^{r}_{\ast j},
\end{equation}
\begin{equation}
    \Delta\delta_{j}=\Delta\delta^b_{\rm HIP}+\Delta\mu^b_{\delta {\rm HIP}}(t_j-t_{\rm HIP})+\Delta\delta^{r}_{j},
\end{equation}
where $\Delta\alpha^{b}_{\ast{\rm HIP}}=(\alpha^b_{\rm HIP}-\alpha_{\rm HIP})\,\cos(\Delta\delta^b_{\rm HIP}/2)$, $\Delta\delta^{b}_{\rm HIP}=\delta^b_{\rm HIP}-\delta_{\rm HIP}$, $\Delta\mu^b_{\alpha {\rm HIP}}=\mu^b_{\alpha {\rm HIP}}-\mu_{\alpha {\rm HIP}}$, and $\Delta\mu^b_{\delta {\rm HIP}}=\mu^b_{\delta {\rm HIP}}-\mu_{\delta {\rm HIP}}$. Therefore, the abscissae of Hipparcos is given by
\begin{equation}
    \hat{\xi_j} = \Delta\alpha_{\ast{j}}\,{\rm cos}\,\psi_j+\Delta\delta_{j}\,{\rm sin}\,\psi_j+\Delta\varpi^{b}_{\rm HIP}f^{\rm AL}_j,~ 
\end{equation}
where $\Delta\varpi^{b}_{\rm HIP}=\varpi^{b}_{\rm HIP}-\varpi_{\rm HIP}$. The above three formulae are slightly different from those of Gaia. We additionally take into account the difference between Hippacos catalog astrometry and the astrometry propagated from the GDR3 epoch to the Hipparcos epoch. This difference includes the long-term position variation and PMa between the Hipparcos and Gaia measurements.
In addition, the scan angle in the new Hipparcos IAD is complementary to the Gaia scan angle \citep{vanLeeuwen2007}. Finally, we can calculate the likelihood for Hipparcos IAD ($\xi_j$) using
\begin{equation}
  \mathcal{L}_{\rm hip}=\prod\limits_{j=1}^{N_{\rm IAD}}\frac{1}{\sqrt{2\pi(\sigma_{j}^2+J_{\rm hip}^2)}}{\rm exp}\left(-\frac{(\hat{\xi}_j-\xi_j)^2}{2(\sigma_{j}^2+J_{\rm hip}^2)}    \right)~,
\end{equation}
where $N_{\rm IAD}$ is the total number of Hipparcos IAD, $\sigma_j$ is the uncertainty of the individual measurement and $J_{\rm hip}$ is the jitter term.

\subsection{Prior and MCMC setting}
We adopt uniform priors for most fitting parameters (see Table~\ref{Tab:result}). Specifically, the stellar mass is assigned a Gaussian prior using the value listed in Table~\ref{Tab:stellar}. With total likelihood $\mathcal{L}$ (depends on any combinations of $\mathcal{L}_{\rm ETV}$, $\mathcal{L}_{\rm RV}$, $\mathcal{L}_{\rm DR2}$, $\mathcal{L}_{\rm DR3}$, $\mathcal{L}_{\rm DR4}$ and $\mathcal{L}_{\rm hip}$), we finally derive the orbital solution by sampling the posterior via the parallel-tempering Markov Chain Monte Carlo (MCMC) sampler \texttt{ptemcee} \citep{Vousden2016}. The RV part is defined in our previous work and will not be repeated here.
We employ 30 temperatures, 100 walkers, and 80,000 steps per chain to generate posterior distributions for all the fitting parameters, with the first 40,000 steps being discarded as burn-in.

\begin{figure}[ht]
    \centering
    \includegraphics[width=0.45\textwidth]{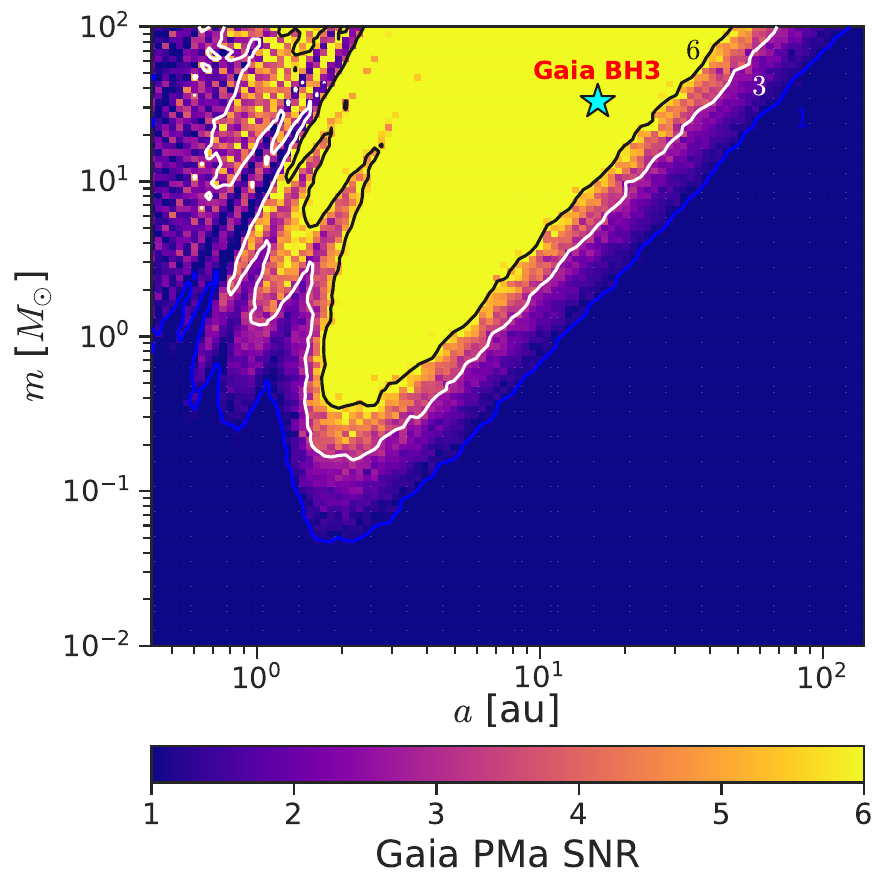}\caption{Gaia PMa SNR map across the $m-a$ space for Gaia BH3 systems. The blue, white and black contour lines respectively correspond to SNR$=$1, 3 and 6. The cyan stars denote the currently observed values of Gaia BH3 in $m-a$ space. It is evident that the theoretical PMa of the system induced by the unseen companion is significant. Because the orbital elements are sampled uniformly, the SNR map reflects only the average behaviour and may differ for any one specific orbital configuration. 
\label{fig:pms_bh}}
\end{figure}

\begin{figure*}[ht]
    \centering
    \includegraphics[width=0.9\textwidth]{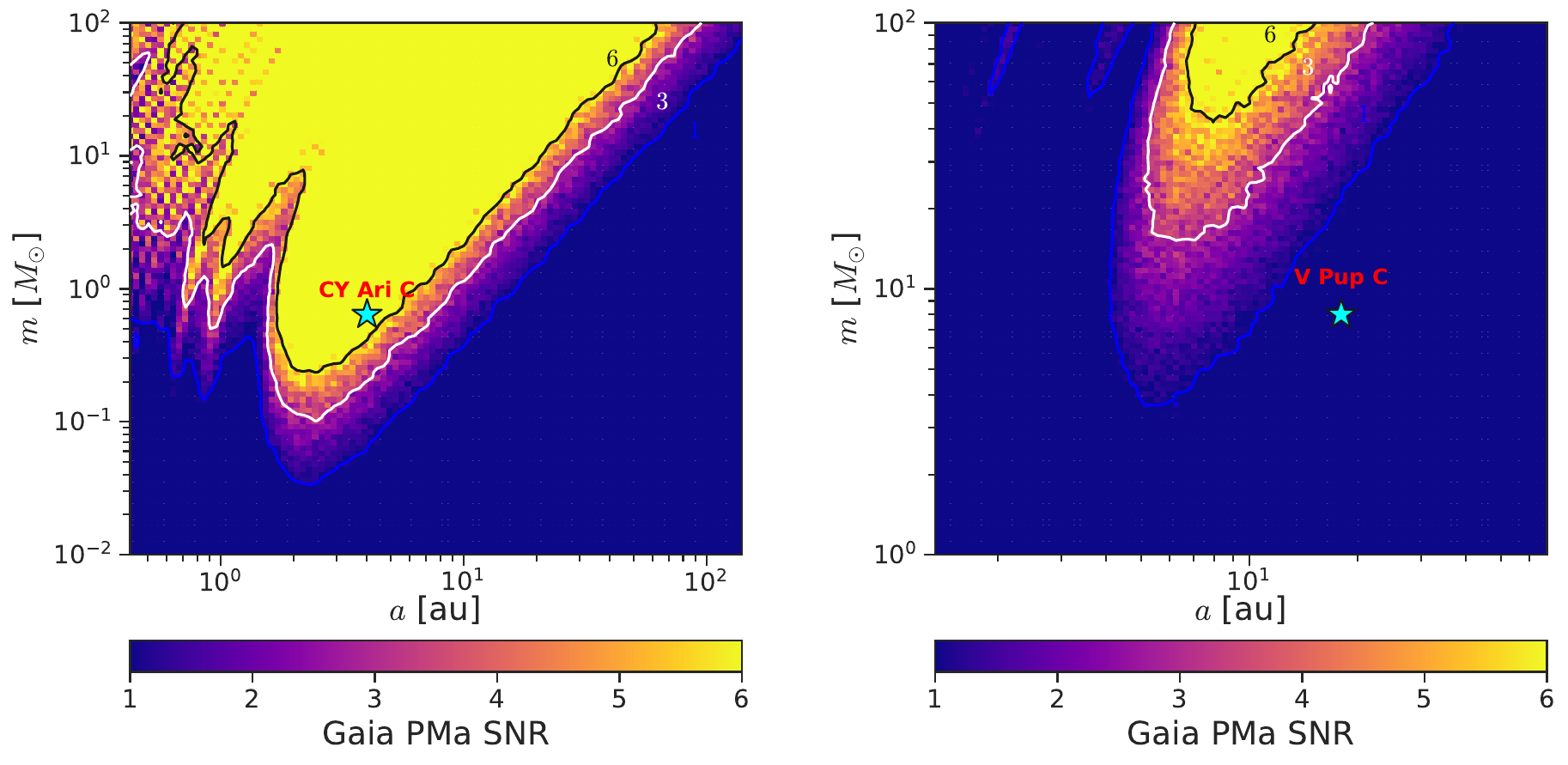}\caption{Gaia PMa significance map across the $m-a$ space for CY Ari and V Pup systems. Symbols are similar to Figure~\ref{fig:pms_bh}. Compared with the CY Ari system, the PMa induced by V Pup C is weak, owing to the large uncertainties in the Gaia astrometry.
\label{fig:pms}}
\end{figure*}

\section{Astrometric signature indicators}

The renormalised unit weight error (RUWE) is commonly used both to assess the quality of an astrometric solution and to set upper or lower limits on the orbit or mass of an unresolved companion (e.g., \citealt{Kiefer2024arXiv}). It is widely accepted that a value of $\rm{RUWE}<1.4$ indicates a good single-star fit, while $\rm{RUWE}>1.4$ implies the unresolved multiplicity of a star~\citep{Lindegren2018,GaiaCollaboration2023_ruweA&A}. For V Pup, CY Ari and Gaia BH3 systems, we find $\rm{RUWE}=2.41$, $1.88$ and $3.41$, respectively, consistent with the multiplicity revealed by recent studies.

In addition, PMa between different catalogs can also be regarded as an additional indicator of the presence of unsolved companions (e.g., \citealt{Kervella2022}). We define the signal-to-noise ratio (SNR) of PMa between two catalogs as
\begin{equation}
    {\rm SNR} = \sqrt{(\Delta \mu)\,\Sigma^{-1}\,(\Delta \mu)^T},
\end{equation}
where $\Delta \mu=[\Delta\mu_\alpha, \Delta\mu_\delta]$ is the vector of the proper motion difference and $\Sigma$ is the relevant covariance matrix.

For targets with both Hipparcos and Gaia measurements (e.g., V Pup), the PMa relative to a long-term proper motion $\mu_{\rm HG}$ (the position difference between Hipparcos and Gaia divided by the $\sim25$ yr baseline) can be calculated from the Hipparcos-Gaia Catalog of Accelerations (HGCA; \citealt{Brandt2021}). As shown in Table~\ref{tab:hgca}, the PMa of V Pup have an SNR of 4.6 at the Hipparcos epoch, and 4.4 at the Gaia epoch. 

\begin{table}[ht]
\centering
\caption{Long-term and short-term PMa of three systems}\label{tab:hgca}
\begin{tabular}{lcccc}
\hline \hline
Target&Catalog & $\Delta \mu_\alpha$& $\Delta \mu_\delta$& SNR \\
&&mas yr$^{-1}$&mas yr$^{-1}$&\\
\hline
V Pup&Hipparcos&$-1.07\pm0.45$&$-1.43 \pm0.37$&4.6\\
&GDR3&$0.26\pm0.35$&$1.52 \pm0.35$&4.4\\
CY Ari&GDR23&$0.59\pm0.10$&$0.36\pm0.08$&8.1\\
Gaia BH3&GDR23&$0.02\pm0.08$&$1.49\pm0.08$&19.0\\
\hline
\end{tabular}
\end{table}

For most targets with only Gaia astrometry, we instead calculate the PMa between GDR2 and GDR3 as an indicator. For CY Ari, we find the proper motion differences are $\Delta\mu_{\alpha}=0.59\pm0.10\,{\rm mas\,yr^{-1}}$ and $\Delta\mu_{\delta}=0.36\pm0.08\,{\rm mas\,yr^{-1}}$ between GDR2 and GDR3, corresponding to SNR=8.1. For Gaia BH3 system, we obtain $\Delta\mu_{\alpha}=0.02\pm0.08\,{\rm mas\,yr^{-1}}$ and $\Delta\mu_{\delta}=1.49\pm0.08\,{\rm mas\,yr^{-1}}$ (SNR=19.0). Both systems show significant short-term PMa. 

To quantify the PMa between GDR23 of a target, we simulate Gaia epoch data with GOST across a range of orbital configurations. The semi-major axis $a$ is uniformly sampled from $0.4\sim34$\,au, and the companion mass $m$ from $0.01\sim100\,M_{\odot}$ ($100\times100$ grids with 25 orbits for each grid). All other elements, except eccentricity, which is fixed at 0, are taken from the same uniform distributions as their priors (see Table~\ref{Tab:result}). For simplicity, we adopt a single companion model and, at the GDR3 epoch, place the barycentre at the origin with the GDR3 proper motion values.
For each orbit, we fit the simulated GDR2 and GDR3 epoch data with a four-parameter astrometric model ($\varpi$ fixed to the GDR3 value) to obtain the synthetic proper motions (Equation \ref{5-p_ast}) and then compute the corresponding PMa SNR. For each grid, we calculate the mean SNR of 25 orbits.

Figure~\ref{fig:pms_bh} and \ref{fig:pms} show the PMa SNR maps from our simulation. For Gaia BH3 and CY Ari systems, the observed PMa SNR is in good agreement with the prediction of the simulation, supporting the idea that the astrometric signature induced by their outer companions is absorbed within GDR2 and GDR3. In contrast, V Pup shows weak short-term PMa due to the large uncertainty of Gaia astrometry (it is too bright, G=4.46 mag), but its long-term PMa between Hipparcos and GDR3 is still evident.
Consequently, the implementation of Hipparcos astrometry and/or multiple Gaia data releases makes it possible to constrain systems that host massive dark companions.

\section{Orbit analysis and results}
\subsection{Verifying the joint analysis with Gaia BH3}
Gaia BH3 was discovered through a preliminary astrometric binary solution of Gaia epoch data. Both the Gaia RVs and the epoch astrometry confirm the presence of a $32.70\,M_\odot$ dormant black hole on a 11.6-yr orbit around a metal-poor star \citep{GaiaBH3_2024}. We adopt this target to verify the robustness of our joint model (mainly GDR23 model) for two considerations: (1) the RVs and the LTTE-only ETVs provide the same orbital information for a Keplerian motion; (2) Gaia BH3 is currently the only source whose astrometric epoch data have been published. Given that the Hipparcos–Gaia model has been extensively validated for nearby stars hosting massive, long-period exoplanets (e.g.,\citealt{Kervella2022, Feng2023MNRAS}, also see the publicly available package \texttt{orvara}, \citealt{Brandt2021_orvara}), we omit its revalidation here.
We conduct a comparative analysis between the RV+GDR23 and RV+GDR4 models. The RVs and astrometric epoch data are taken from \citet{GaiaBH3_2024}, and the GDR23 catalog astrometry is retrieved from Gaia archive (see Table~\ref{Tab:stellar}).

\begin{figure*}[ht!]
    \centering
    \includegraphics[width=0.9\textwidth]{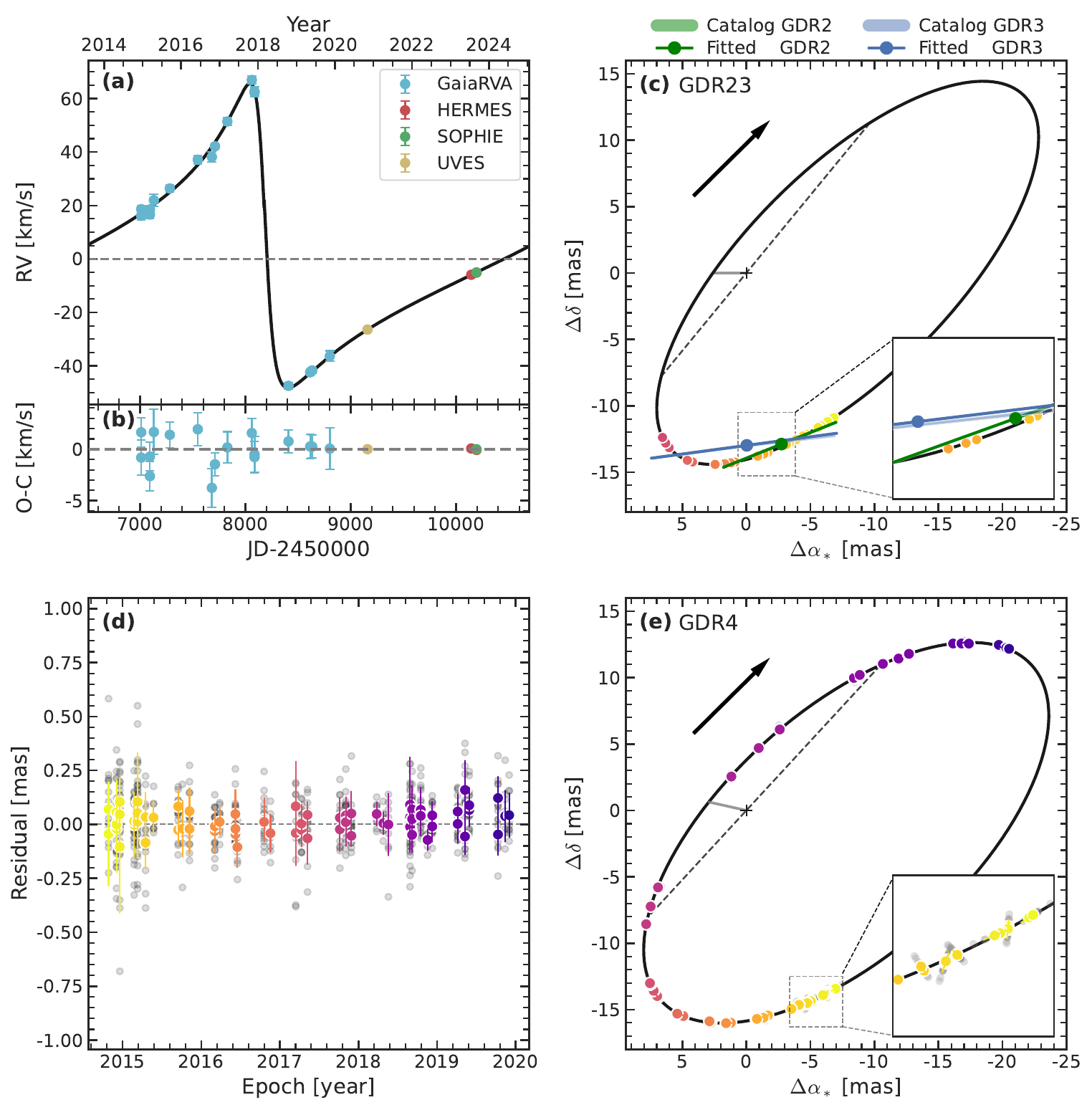}\caption{Comparison of RV+GDR23 and RV+GDR4 model for Gaia BH3. \textbf{Panel (a):} the best-fit orbit to RVs from RV+GDR4 model. Because the RV part of these two model is nearly identical, we only show one result to avoid repetition. \textbf{Panel (b):} RV residuals between best-fit solution and observations. \textbf{Panel (c):} the best-fit astrometric orbit of the star (or photocentre) in the sky-projected plane from RV+GDR23 model. The central plus symbol marks the barycenter and the gray line connects it with the position of the periastron. The thich dotted line shows the line of nodes, and the arrow indicates the direction of the motion along the orbit. Gaia epoch data simulated with GOST are indicated by colored solid dots. The small panel in the lower right corner is an enlargement of the region of the fitting to GDR23, depicting the best fit to Gaia GOST data and the comparison between the best-fit and catalog astrometry (positions and proper motions) at GDR2 and GDR3 reference epochs. The shaded regions represent the uncertainty of catalog positions and proper motions after removing the motion of the BOS. The two segments and their center dots (green and blue) represent the best-fit proper motion and position offsets induced by the black hole at GDR2 and GDR3, respectively. \textbf{Panel (d):} the residuals of the along-scan (AL) astrometric measurements for RV+GDR4 solution. The gray dots denote the original data, while the colored dots with error bars show the binned data for each transit. \textbf{Panel (e):} the best-fit astrometric orbit from RV+GDR4 model. The post-fit residuals are projected into the R.A. and decl. axes (gray dots). For panel (c), (d) and (e), the dots of the same color share identical orbital phase.
\label{fig:BH3}}
\end{figure*}

Table~\ref{Tab:result} lists the optimal orbital solutions from the two models, and Figure~\ref{fig:BH3} shows the corresponding fits to RVs, Gaia AL abscissae, and GOST data. The RV+GDR23 solution yields a black-hole mass of $30.4_{-1.7}^{+1.8}\,M_\odot$, consistent within $1.4\sigma$ and $1.2\sigma$ with the RV+GDR4 value ($33.8_{-0.47}^{+0.48}\,M_\odot$) and the literature value ($32.7\pm0.82\,M_\odot$ of \citealt{GaiaBH3_2024}), although the uncertainty is slightly greater. Because the RVs do not yet cover a full orbit and the GDR23 baseline is only half that of GDR4, the period is less tightly constrained, but remains compatible with the other two results. The corner plots of the posterior distribution of the orbital parameters are presented in Appendix~\ref{app:corner}.
Overall, the RV+GDR23 model successfully recovers the orbital parameters and the true property of Gaia BH3 to within an acceptable range, mainly benefiting from the large astrometric signal within the GDR23. Our joint RV+GDR23 model has recently revealed G3425 (2MASS J06170689+2343487) as a wide-binary containing a stellar-mass black hole ($3.6_{-0.5}^{+0.8}\,M_\odot$ \citealt{Wang2024NatAs_G2546}). 

\begin{table*}[!]
\begin{rotatetable*}
\centering
\caption{Parameters for V Pup, CY\,Ari and Gaia BH3 system}\label{Tab:result}
\begin{tabular*}{1.2\textwidth}{@{}@{\extracolsep{\fill}}lcccccc@{}}
\hline \hline
Parameter & V Pup & CY Ari & \multicolumn{3}{c}{Gaia BH3} & Prior${\rm ^a}$ \\
\cline{4-6}
&ETV+HG23&ETV+GDR23& RV+GDR23 & RV+GDR4 & Reference$\rm ^e$ &  \\
\hline
\textbf{Orbital parameter}&\\
Orbital period $P$ (days)&${5126.08}_{-0.94}^{+0.95}$&${1974.5}_{-5.8}^{+6.2}$   &${4615}_{-444}^{+550}$&${4227}_{-87}^{+90}$&$4253.1\pm98.5$&Log-$\mathcal{U}(-1,16)$\\
RV semi-amplitude $K$ (km s$^{-1}$)&${11.353}_{-0.071}^{+0.073}$&${7.87}_{-0.18}^{+0.22}$  &${57.4}_{-1.0}^{+1.2}$&${56.85}_{-0.30}^{+0.30}$&---&$\mathcal{U}(10^{-6},10^{6})$\\
Eccentricity $e$&${0.4587}_{-0.0046}^{+0.0046}$& ${0.526}_{-0.027}^{+0.032}$  &${0.753}_{-0.029}^{+0.028}$&${0.7280}_{-0.0043}^{+0.0044}$&$0.7291\pm0.0048$&$\mathcal{U}(0,1)$\\
Argument of periapsis $\omega_b$ ($^\circ$)$\rm ^b$& ${202.55}_{-0.45}^{+0.46}$& ${85.6}_{-3.0}^{+2.8}$ &${75.9}_{-2.6}^{+2.7}$&${77.79}_{-0.16}^{+0.16}$&$77.34\pm0.76$&$\mathcal{U}(0,360)$\\
Mean anomaly at J2017.5 $M_{0}$ ($^\circ$)&${154.67}_{-0.44}^{+0.44}$& ${342.0}_{-2.7}^{+3.0}$ &${342.5}_{-3.7}^{+2.8}$&${339.55}_{-0.46}^{+0.46}$&---&$\mathcal{U}(0,360)$\\
Longitude of ascending node $\Omega$ ($^\circ$)&${4.6}_{-3.3}^{+5.7}$& ${317}_{-29}^{+22}$  &${142.2}_{-7.0}^{+9.4}$&${136.21}_{-0.12}^{+0.13}$&$136.236\pm0.128$&$\mathcal{U}(0,360)$\\
Inclination $I$ ($^\circ$)$\rm ^c$&${87.9}_{-3.3}^{+3.3}$& ${85.6}_{-6.5}^{+7.8}$  &${106.5}_{-4.5}^{+3.3}$&${110.601}_{-0.093}^{+0.094}$&$110.580\pm0.095$&Cos$I$-$\mathcal{U}(-1,1)$\\
\hline
\textbf{Derived parameter}\\
Orbital period $P$ (yr)&${14.0344}_{-0.0026}^{+0.0026}$& ${5.406}_{-0.016}^{+0.017}$  & ${12.6}_{-1.2}^{+1.5}$&${11.57}_{-0.24}^{+0.25}$&$11.64\pm0.27$&---\\
Semi-major axis $a$ (au)&${17.88}_{-0.15}^{+0.15}$& ${3.906}_{-0.082}^{+0.079}$  &${17.1}_{-1.3}^{+1.5}$&${16.55}_{-0.26}^{+0.26}$&$16.17\pm0.27$&---\\
Periapsis epoch $T_{\rm p}-2400000$ (JD)&${60860.6}_{-6.4}^{+6.4}$& ${58036}_{-16}^{+15}$   &${58163}_{-563}^{+462}$&${58177.00}_{-0.94}^{+0.92}$&$58177.39\pm0.88$&---\\
Companion mass $m$ ($M_{\odot}$)& ${7.73}_{-0.14}^{+0.14}$& ${0.640}_{-0.029}^{+0.029}$   &${30.4}_{-1.7}^{+1.8}$&${33.08}_{-0.47}^{+0.48}$&${32.70}\pm0.82$&---\\
\hline
\textbf{Barycenter parameter}\\
$\alpha*$ offset $\Delta \alpha*$ (mas)$\rm ^d$&${1.09}_{-0.89}^{+1.5}$& ${-1.25}_{-0.27}^{+0.30}$   &${-0.14}_{-0.80}^{+0.78}$&${4.288}_{-0.051}^{+0.052}$&---&$\mathcal{U}(-10^{6},10^{6})$\\
$\delta$ offset $\Delta \delta$ (mas)&${14.86}_{-0.81}^{+0.81}$& ${0.80}_{-0.35}^{+0.31}$  &${-12.9}_{-1.4}^{+1.4}$&${2.401}_{-0.067}^{+0.067}$&---&$\mathcal{U}(-10^{6},10^{6})$\\
Proper motion $\mu_{\alpha*}$ (mas\,yr$^{-1}$)&${-5.53}_{-0.26}^{+0.26}$& ${28.24}_{-0.60}^{+0.78}$   &${-27.43}_{-0.90}^{+0.90}$&${-28.341}_{-0.059}^{+0.058}$&$-28.317\pm0.067$&$\mathcal{U}(-10^{6},10^{6})$\\
Proper motion $\mu_\delta$ (mas\,yr$^{-1}$)&${8.2747}_{-0.25}^{+0.25}$& ${-14.47}_{-1.0}^{+0.41}$   &${-154.4}_{-1.1}^{+1.1}$&${-155.19}_{-0.10}^{+0.10}$&$-155.221\pm0.111$&$\mathcal{U}(-10^{6},10^{6})$\\
Parallax $\varpi$ (mas)&${2.40}_{-0.23}^{+0.23}$& ${3.052}_{-0.034}^{+0.033}$   &${1.675}_{-0.078}^{+0.078}$&${1.6844}_{-0.0080}^{+0.0080}$&$1.6933\pm0.0164$&$\mathcal{U}(-10^{6},10^{6})$\\
\hline
\textbf{Other parameter}\\
Jitter for Hipparcos epoch data $J_{\rm hip}$ (mas)&${1.69}_{-0.15}^{+0.16}$& ---  &---&---&---&$\mathcal{U}(0,10^{6})$\\
Jitter for Gaia epoch data $J_{\rm gaia}$ (mas)&---& ---  &---&${0.032}_{-0.012}^{+0.010}$&---&$\mathcal{U}(0,10^{6})$\\
Error inflation factor $S_{\rm gaia}$&${1.103}_{-0.065}^{+0.080}$& ${1.087}_{-0.057}^{+0.078}$   &${1.091}_{-0.060}^{+0.079}$&---&---&$\mathcal{N}(1,0.1)$\\
Corrected reference epoch $\Delta T_0$ (days)&${-1.579(0.021)\times10^{-2}}$&${6.98(0.25)\times10^{-3}}$&---&---&---&$\mathcal{U}(-10^{6},10^{6})$\\
Corrected binary period $\Delta P_0$ (days)&${-1.705(0.034)\times10^{-6}}$&${-1.29(0.06)\times10^{-6}}$&---&---&---&$\mathcal{U}(-10^{6},10^{6})$\\
Rate of period change $\dot{P}$ ($s\cdot s^{-1}$)&${6.345(0.058)\times10^{-10}}$& ${6.29(0.20)\times10^{-10}}$&---&---&---&$\mathcal{U}(-10^{6},10^{6})$\\
\hline
\multicolumn{7}{l}{$\rm ^a$ Log-$\mathcal{U}(a, b)$ is the logarithmic uniform distribution between $a$ and $b$, Cos$i$-$\mathcal{U}(a, b)$ is the cosine uniform distribution of $I$}\\
\multicolumn{7}{l}{between $a$ and $b$, and $\mathcal{N}(a, b)$ is the Gaussian distribution with mean $a$ and stardard deviation $b$.}\\
\multicolumn{7}{l}{$\rm ^b$ The argument of periastron of the stellar reflex motion, differing by $\pi$ with companion orbit, i.e., $\omega_{\rm p}=\omega+\pi$.}\\
\multicolumn{7}{l}{$\rm ^c$ For RV+GDR23 model of Gaia BH3 system, we add an additional bound prior ($i>90^\circ$) for the inclination.}\\
\multicolumn{7}{l}{$\rm ^d$ The reference position of Gaia BH3 for GDR4 is $\alpha_0=294^\circ.82784900557243$ and $\delta_0=14^\circ.930918410309376$, instead of GDR3 values.}\\
\multicolumn{7}{l}{$\rm ^e$ Reference of Gaia BH3 is from \citet{GaiaBH3_2024}.}
\end{tabular*}
\end{rotatetable*}
\end{table*}

\subsection{Joint analysis of ETV and astrometry}

\subsubsection{V Pup}
V Pup is a young (5 Myr), bright (G = 4.46 mag) multiple system composed of a semidetached eclipsing binary with two massive B-type components (e.g., \citealt{Schneider1979AJ, Andersen1983A&A, Bell1987MNRAS}) and a distant, stellar-mass black-hole candidate \citep{Qian2008ApJ}. Previous studies have identified a weak X-ray source ($\sim0.5$ arcsec, \citealt{Giacconi1974ApJS,Bahcall1975MNRAS,Maccarone2009MNRAS}) and an H$\mathrm{II}$ region ($\sim3.5$ arcmin, \citealt{York1976ApJ,Koch1981PASP}) in the vicinity of the binary. 
\citet{Qian2008ApJ} attributed the periodic O–C variation of V Pup to a third-body orbit with a period of 5.47 yr and a minimum mass of $10.4\,M_\odot$, implying a stellar-mass black hole.
However, \citet{Budding2021MNRAS} found that their updated O-C measurements deviate significantly from the model of \citet{Qian2001MNRAS}.

We include the ETV data of \citet{Budding2021MNRAS}\footnote{\url{https://www.variablestarssouth.org/community/articles/budding-love-blackford-banks-and-rhodes-2020-absolute-parameters-of-young-stars-v-puppis}}, as well as new measurements of the minimum time from the TESS mission \citep{Ricker2015}, BRITE collaboration\footnote{\url{https://brite.camk.edu.pl/pub/index.html}} \citep{Popowicz2017A&A_BRITE} and the O-C gateway\footnote{\url{https://var.astro.cz/en}} for the joint fitting. To determine the time of minima in the TESS light curve (PDC\_SAP flux; \citealt{Smith2012PASP,Stumpe2012PASP,Stumpe2014PASP}), we first phased and averaged all the data into a single orbital period. We then fitted the region around each minimum with a high-order polynomial to create a template. This template was subsequently cross-correlated with every individual minimum to measure the phase shift; from the shift we computed the corresponding time of minima (largely following \citealt{Borkovits2015MNRAS}). Because the cadence of the BRITE light curve is not as consecutive as TESS, we binned multiple light curves (e.g., \citealt{LiKai2021ApJ}) to derive the mean minimum times. 

We recalculate the O-C values using the following ephemeris of eclipse minima:
\begin{equation}
{\rm Min.}I\,{\rm (BJD)}= 2445367.621616+ 1.454492\times E.
\end{equation}
The time scale of ETV is usually given in [BJD, TDB\footnote{BJD: Barycentric Julian Date; TDB: Barycentric Dynamical Time. HJD (Heliocentric Julian Date) to BJD: \url{https://astroutils.astronomy.osu.edu/time/hjd2bjd.html}}], whereas the Gaia GOST and epoch data are provided in [BJD, TCB\footnote{Barycentric Coordinate Time.}]. It is better to correct for this small difference, although it has little effect on the final orbital accuracy.

\begin{figure}[ht]
    \centering
    \includegraphics[width=0.45\textwidth]{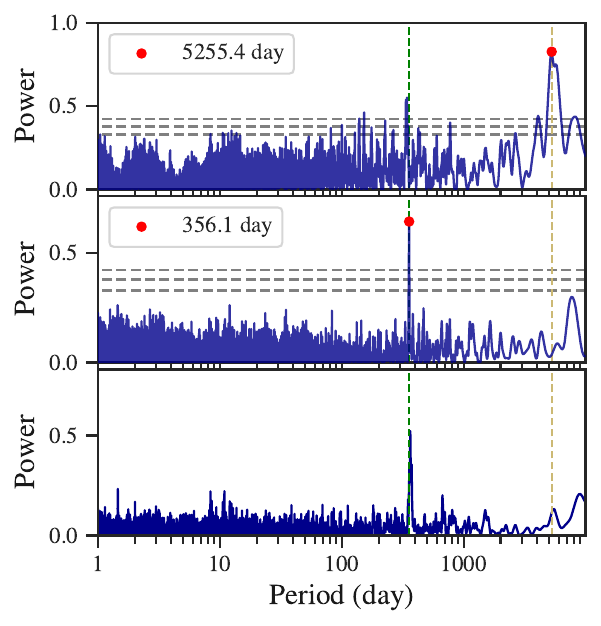}\caption{Generalized Lomb-Scargle (GLS) periodograms for V Pup ETV. \textbf{Upper panel:} the periodogram for O-C data with the parabolic trend removed. A periodic signal at 5255 day is significant. \textbf{Middle panel:} the periodogram for O-C residuals. A signal near the sample window emerges (365 day). The horizontal grey lines, top to bottom, indicate the 0.001, 0.01, 0.1 False Alarm Probability (FAP) levels, respectively. \textbf{Bottom panel:} window function.
\label{fig:period}}
\end{figure}

\begin{figure*}[ht!]
    \centering
    \includegraphics[width=0.9\textwidth]{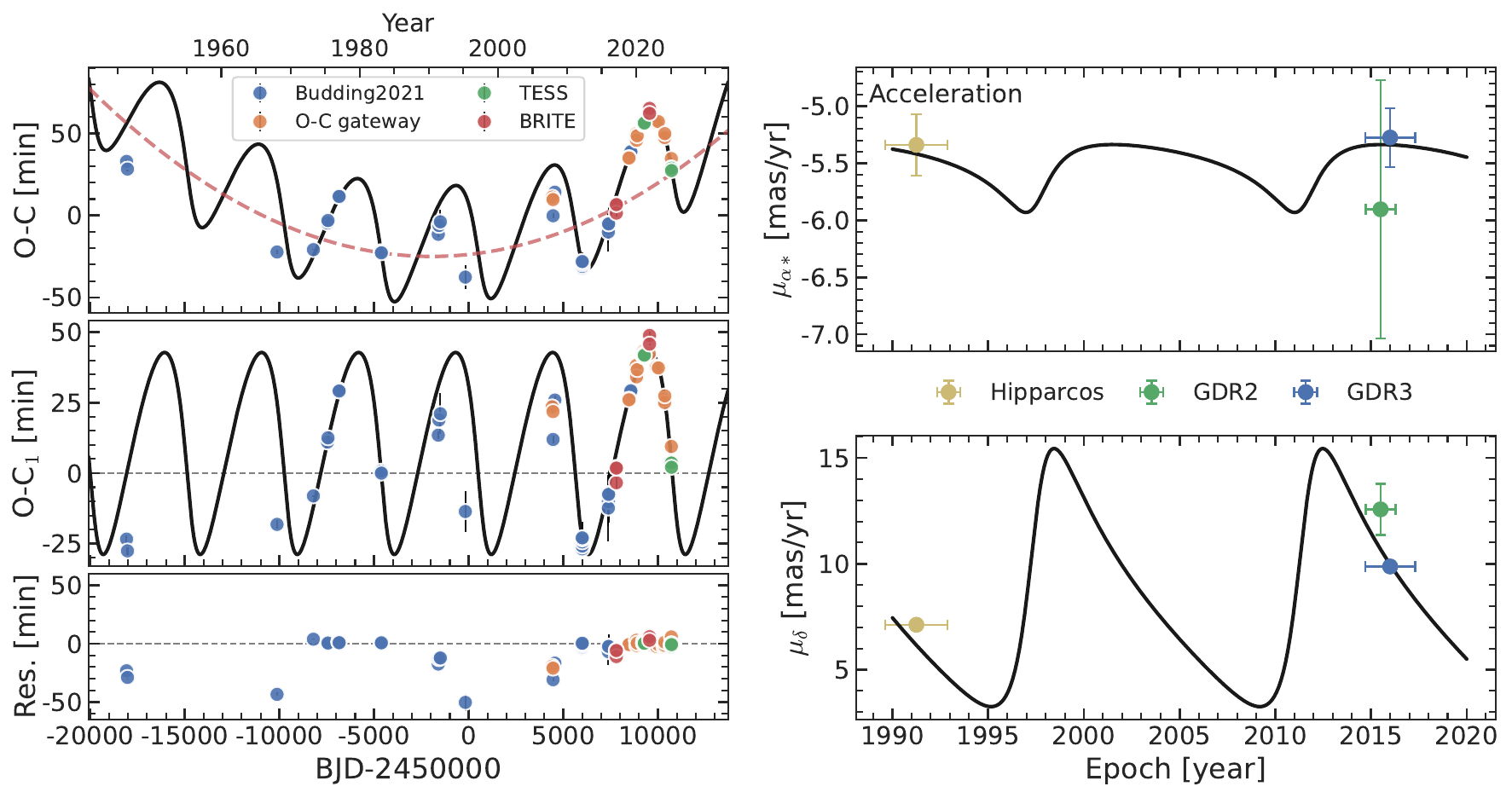}\caption{Best-fit orbit to ETVs and Hip-Gaia astrometry of V Pup from our joint fitting. \textbf{Top-left panel:} the thick black line is the best-fit curve to ETVs, and the red dashed line is the quadratic term. \textbf{Middle-left panel:} the ETV curve (pure LTTE) after correcting the quadratic term. \textbf{Bottom-left panel:} the final residuals.  \textbf{Right panel:} the absolute astrometric acceleration in right ascension and declination. The horizontal error bar of each data point denotes the temporal baseline of each catalog.
\label{fig:Vpup}}
\end{figure*}

Figure~\ref{fig:period} shows the Generalized Lomb-Scargle (GLS; \citealt{Zechmeister2009}) periodograms for the ETV of V Pup. A significant periodic signal of 5255 day (14.39 yr) emerges, which is longer than the one (5.47 yr) found by \citet{Qian2008ApJ}. Our joint analysis reveals that the outer companion V Pup C has a period of $P_C={14.0344}_{-0.0026}^{+0.0026}$ yr, an eccentricity of $e_C={0.4587}_{-0.0046}^{+0.0046}$, a true mass of $M_C={7.73}_{-0.14}^{+0.15}\,M_\odot$ and an inclination of $I_C={{87.9}_{-3.3}^{+3.3}}^\circ$ (see Table~\ref{Tab:result}). The large mass and absence of spectral lines for the third body, as identified in previous studies (e.g.,\citealt{Andersen1983A&A,Qian2008ApJ, Budding2021MNRAS}), suggest that V Pup C may be a stellar-mass black hole. 
Figure~\ref{fig:Vpup} depicts the best-fit solution to ETVs and Hipparcos-Gaia astrometry. We find that the O-C data collected after 2010 generally exhibit a smaller residual scatter than those obtained before 2010 (bottom-left panel). The reason remains unclear. Either technical factors (such as heterogeneous procedures for extracting the times of minimum or inconsistent barycentric time corrections) or physical mechanisms (e.g., contamination from background star or other, as yet unidentified, stellar activities) could bias or even modulate the observed cyclic signal.

\subsubsection{CY Ari}

We use the ETV data compiled by \citet{Xu_CYAri_2025ApJ} and the GDR23 astrometry from the Gaia archive for our joint fit. 
Spectroscopic and photometric analyses by \citet{Xu_CYAri_2025ApJ} indicate that the outer companion of CY Ari is probably a low-luminosity star, most likely a white dwarf. Under this scenario (dark companion), the system photocentre is dominated by the light of the primary star of the inner pair ($M_A=1.1\,M_\odot$ and $M_B=0.3\,M_\odot$). The ETV-only fit gives a minimum mass of $M_C\,{\rm sin}\,I_C=0.64\,M_\odot$ for the third body. The combined analysis of ETV and GDR23 yields a period of $P_C=5.406_{-0.016}^{+0.017}$ yr, an eccentricity of $e_C=0.526_{-0.027}^{+0.032}$, a true mass of $M_C=0.640_{-0.029}^{+0.029}\,M_\odot$ and an inclination of $I_C={85.6_{-6.5}^{+7.8}}^\circ$. Our period is fully consistent with the $P_C=5.39\pm0.03$ yr reported by \citet{Xu_CYAri_2025ApJ}, while our eccentricity is slightly higher than $0.42\pm0.06$ of theirs, but still within $1.6\sigma$. The nearly edge-on inclination suggests that the orbit of the outer companion is probably aligned with the inner EB, although the ascending node $\Omega$ of the binary orbit remains unknown.   

If the outer companion were a low-mass M-type star with $M_C=0.64\,M_\odot$, its predicted apparent G-band magnitude would be $\sim17.5$ mag, well below Gaia's detection limit. Therefore, this outer companion would contribute only a small fraction of light to the system photocentre. We repeated the fit using the M–G relation \citep{mamajek2013ApJS} for this luminous companion, but the resulting orbital parameters differ only marginally from those obtained under the dark companion assumption. The solution is omitted here to avoid redundancy. Note that the most reliable way to establish the photocentre is to provide the flux ratio between the outer and inner components (see Equation~\ref{equ:photo}).

\begin{figure*}[ht!]
    \centering
    \includegraphics[width=0.9\textwidth]{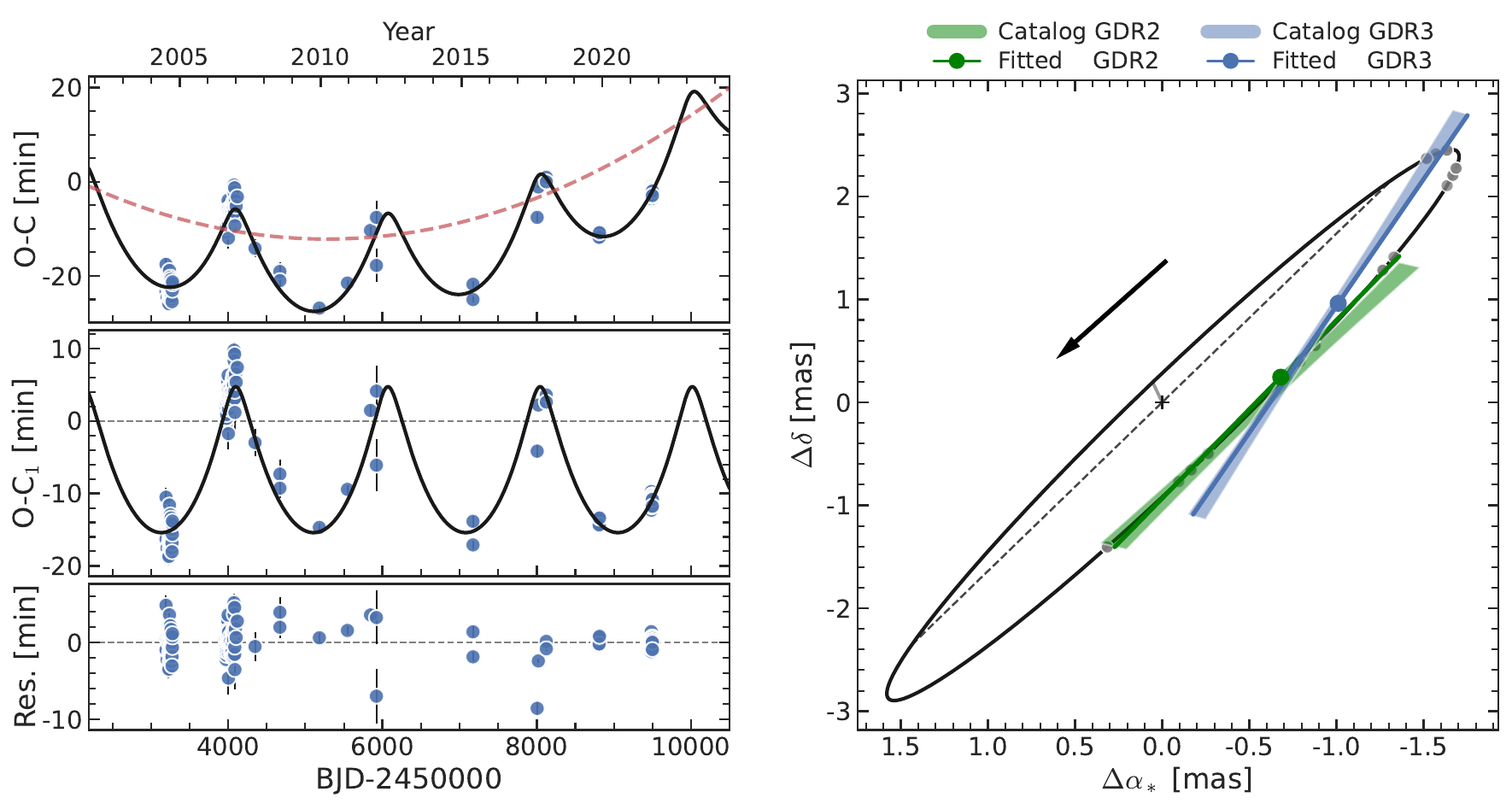}\caption{Best-fit orbit to ETVs and GDR23 of CY Ari from our joint fitting. \textbf{Top-left panel:} the thick black line is the best-fit curve to ETVs, and the red dashed line is the quadratic term. \textbf{Middle-left panel:} the ETV curve (pure LTTE) after correcting the quadratic term. \textbf{Bottom-left panel:} the final residuals. \textbf{Right panel:} the best-fit astrometric orbit of the star in the sky-projected plane, after a subtraction of parallax and system's barycentric motion. Symbols are the same as Panel (c) of Figure~\ref{fig:BH3}.
\label{fig:CYAri}}
\end{figure*}

\section{Discussion and conclusion} 
In this work, we present a method that combines ETV, Hipparcos and/or Gaia astrometry to detect and characterize dark companions around EBs. The method allows us to break the $M\,{\sin}\,I$ degeneracy inherent in LTTE-only analysis and thus to determine both the true mass and the inclination of the tertiary. We restrict our method to close-in EBs orbited by wide companions (e.g., systems like V Pup and CY Ari); for these systems the LTTE is assumed to be the dominant origin of any periodic ETV signal. When the gravitational interaction between inner and outer orbits is no longer negligible, additional dynamical time-delay terms must be included (e.g., \citealt{Borkovits2015MNRAS}). If the outer body is luminous and its light shifts the system photocentre, we recommend using the flux ratio between the outer and inner bodies instead of the M-G relation for the correction. The latter is model-dependent and might introduce additional degeneracy between inclination and companion mass.  

We do not confirm the 5.47-year periodic signal of V Pup reported by \citet{Qian2008ApJ}, instead we find a more long period of ${14.0344}_{-0.0026}^{+0.0026}$ yr. 
The orbital parameters and the mass of this black hole candidate are refined to $a_C={17.88}_{-0.15}^{+0.15}$\,au, $e_C={0.4587}_{-0.0046}^{+0.0046}$ and $M_C={7.73}_{-0.14}^{+0.15}\,M_\odot$. 
\citet{Qian2008ApJ} proposed that the faint X-ray emission might arise from accretion of the wind of the massive binary onto the black hole. However, the relatively large separation ($\sim18$\,au) might make it difficult to occur, so the observed X-ray emission most likely originates from mass transfer between the two B-type components \citep{Maccarone2009MNRAS}. 
The mass of this candidate is comparable to that of the secondary star ($M_B=7.3\,M_\odot$) of the binary. If the candidate is a B-type main-sequence star, it could explain the third-light contribution reported in previous studies (e.g.,\citealt{Budding2021MNRAS}), but would conflict with the absence of any spectroscopic signature. 
If the tertiary is a black hole, the supernova that produced it could also explain the enhanced abundance of heavy elements observed in the surrounding H$\mathrm{II}$ region (e.g.,\citealt{Qian2008ApJ}) near the V Pup system. 
Therefore, a definitive conclusion is not yet possible. We recommend collecting additional spectroscopic, photometric, and astrometric data to establish the true nature of this candidate.

It should be noted that the Applegate mechanism \citep{Applegate1992ApJ} has also been proposed to explain the periodic variations observed in the O-C diagram. In this picture, magnetic activity redistributes angular momentum inside the active star, altering its gravitational quadrupole moment and thereby modulating the orbital period of the close binary. Because the mechanism requires the transfer of angular momentum from the stellar interior to exterior shell, its viability is usually assessed by comparing the energy required for this redistribution with the star's total available energy. For CY Ari, the theoretical calculation of \citet{Xu_CYAri_2025ApJ} suggests that the Applegate process would need several tens of times more energy than the star can supply. Moreover, the significant Gaia PMa between GDR2 and GDR3 also reinforces the hypothesis that a third body is indeed present. Therefore, we conclude that the additional information from Gaia astrometry is crucial for confirming the nature of candidate companions to EBs. However, our method cannot completely rule out a minor contribution from the Applegate mechanism.

One might concern that the eclipse of the inner binary could bias astrometric measurements, because the observation of satellites will sample different orbital phases and hence different flux levels of the binary. For CY Ari (G=11.76 mag), the magnitude variation of eclipse in the V band is $~0.5$,mag, roughly equivalent to the difference in the Gaia G band. According to Figure 3 of \citet{Holl2023A&A}, the median CCD AL abscissa uncertainty is nearly constant ($\sim0.15$ mas) for sources with $9 < G < 13$ mag, so the astrometric precision for CY Ari is not noticeably degraded. Fainter sources could lose some precision, but this effect is largely averaged out over Gaia’s baseline, which is much longer than the binary period. The median uncertainty for the bright source V Pup is slightly larger ($\sim0.35$ mas), but still lower than the astrometric signal of the target ($a_\star\sim19$\,mas). Therefore, Gaia DR4 will provide vital constraints for the tertiary companion of V Pup.    

Overall, our method provides an opportunity to constrain the orbit and mass of many kinds of dark companions to EBs, such as circumbinary planets, brown dwarfs, white dwarfs, neutron stars and black holes, and also enables population-level studies of the formation and evolution of these hierarchical triple systems.

\begin{acknowledgments}
We thank the referee for his constructive suggestions, which have greatly improved this paper. We sincerely thank Fernández-Lajús Eduardo and Anton Paschke for their dedicated efforts in collecting data within the VarAstro framework.
This work is supported by the National Key R\&D Program of China, Nos. 2024YFC2207700 and 2024YFA1611801, and 2023YFA1607901, by the National Natural Science Foundation of China (NSFC) under Grant Nos. 12588202, 12473066, and 12273057, by the Strategic Priority Program of the Chinese Academy of Sciences under Grant No. XDB1160302, by the Shanghai Jiao Tong University 2030 Initiative, by science research grants from the China Manned Space Project, and by the China-Chile Joint Research Fund (CCJRF No. 2205). CCJRF is provided by Chinese Academy of Sciences South America Center for Astronomy (CASSACA) and established by National Astronomical Observatories, Chinese Academy of Sciences (NAOC) and Chilean Astronomy Society (SOCHIAS) to support China-Chile collaborations in astronomy.
\end{acknowledgments}

\begin{contribution}

All authors contributed equally to this work.


\end{contribution}

%
\facilities{Hipparcos, Gaia, TESS, BRITE}

\software{astropy \citep{2013A&A...558A..33A,2018AJ....156..123A,2022ApJ...935..167A},  
          scipy \citep{2020SciPy-NMeth}, matplotlib \citep{Hunter:2007}.}


\appendix
\section{Data availability}
All the data and code used in this work are available on \dataset[Zenodo]{https://doi.org/10.5281/zenodo.17986248}

\section{Additional figures}
\label{app:corner}
\begin{figure*}
    \centering
	\includegraphics[width=\textwidth]{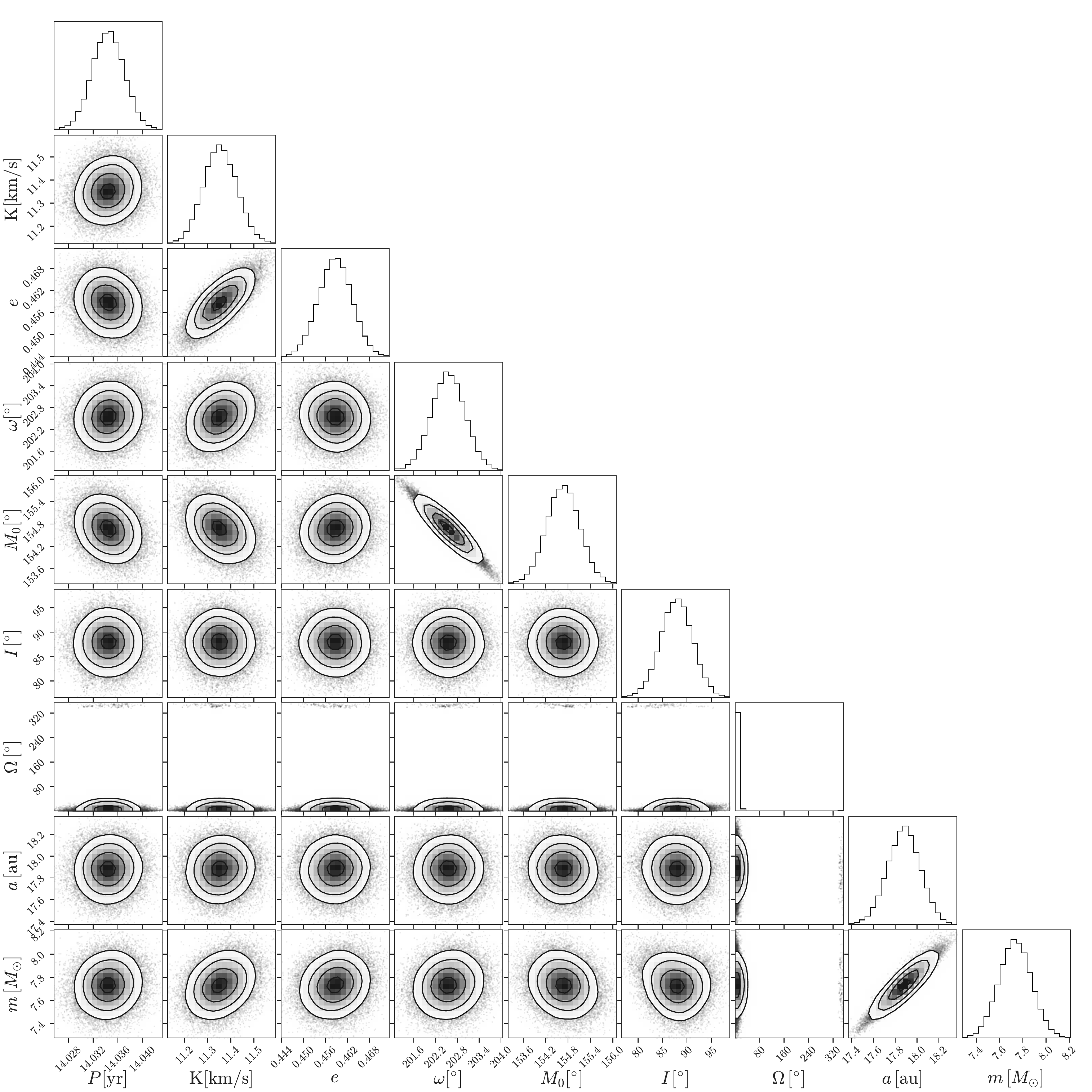}
    \caption{Posterior distributions of the selected orbital parameters for V Pup system. These are the orbital period $P$, the RV amplitude $K$, eccentricity $e$, argument of periapsis $\omega$, Mean anomaly at J2017.5 $M_0$, inclination $I$, Longitude of ascending nod $\Omega$, semi-major axis $a$ and companion mass $m$.}
    \label{fig:corner_hip38957}
\end{figure*}
\begin{figure*}
    \centering
	\includegraphics[width=\textwidth]{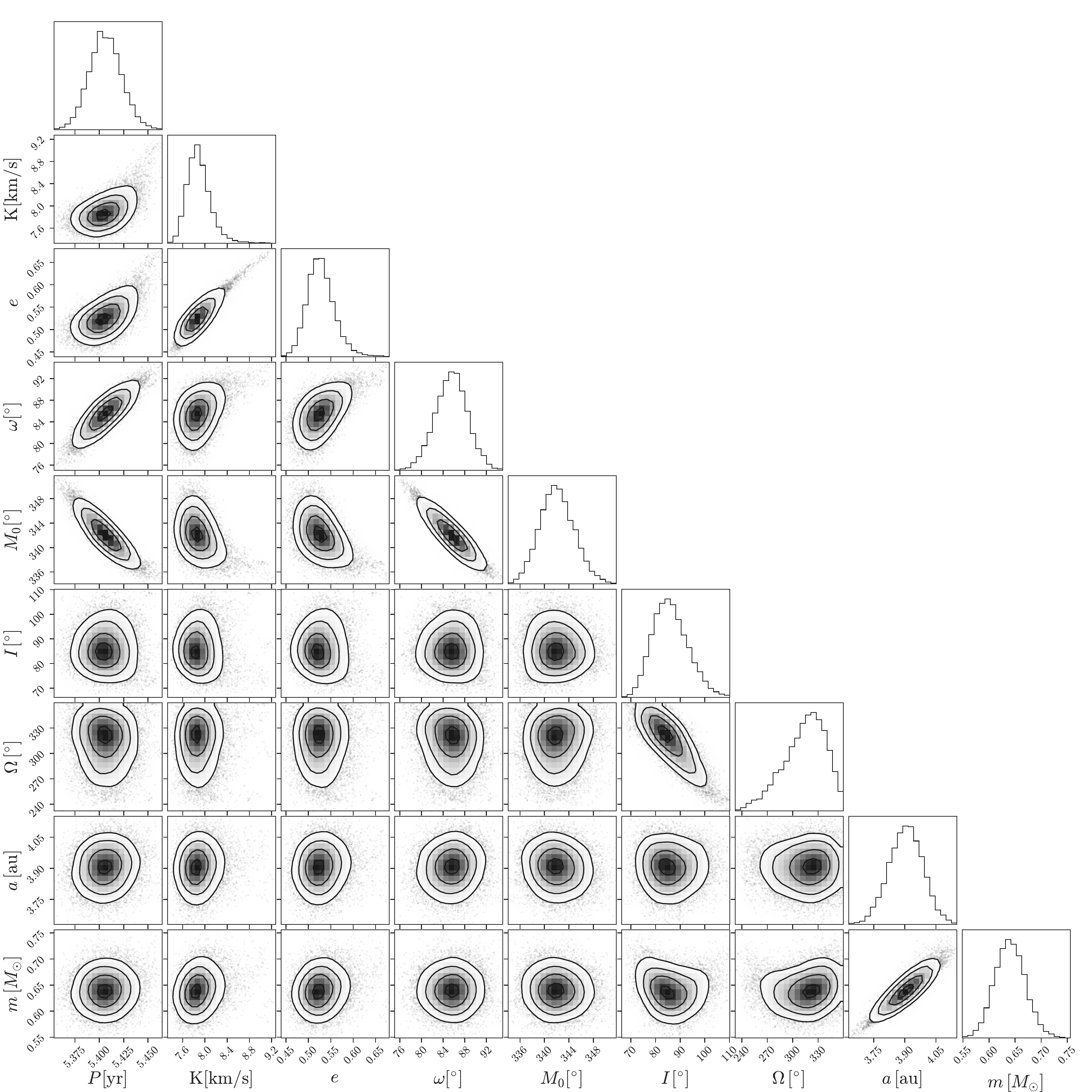}
    \caption{Posterior distributions of the selected orbital parameters for CY Ari system.}
    \label{fig:corner_cyari}
\end{figure*}
\begin{figure*}
    \centering
	\includegraphics[width=\textwidth]{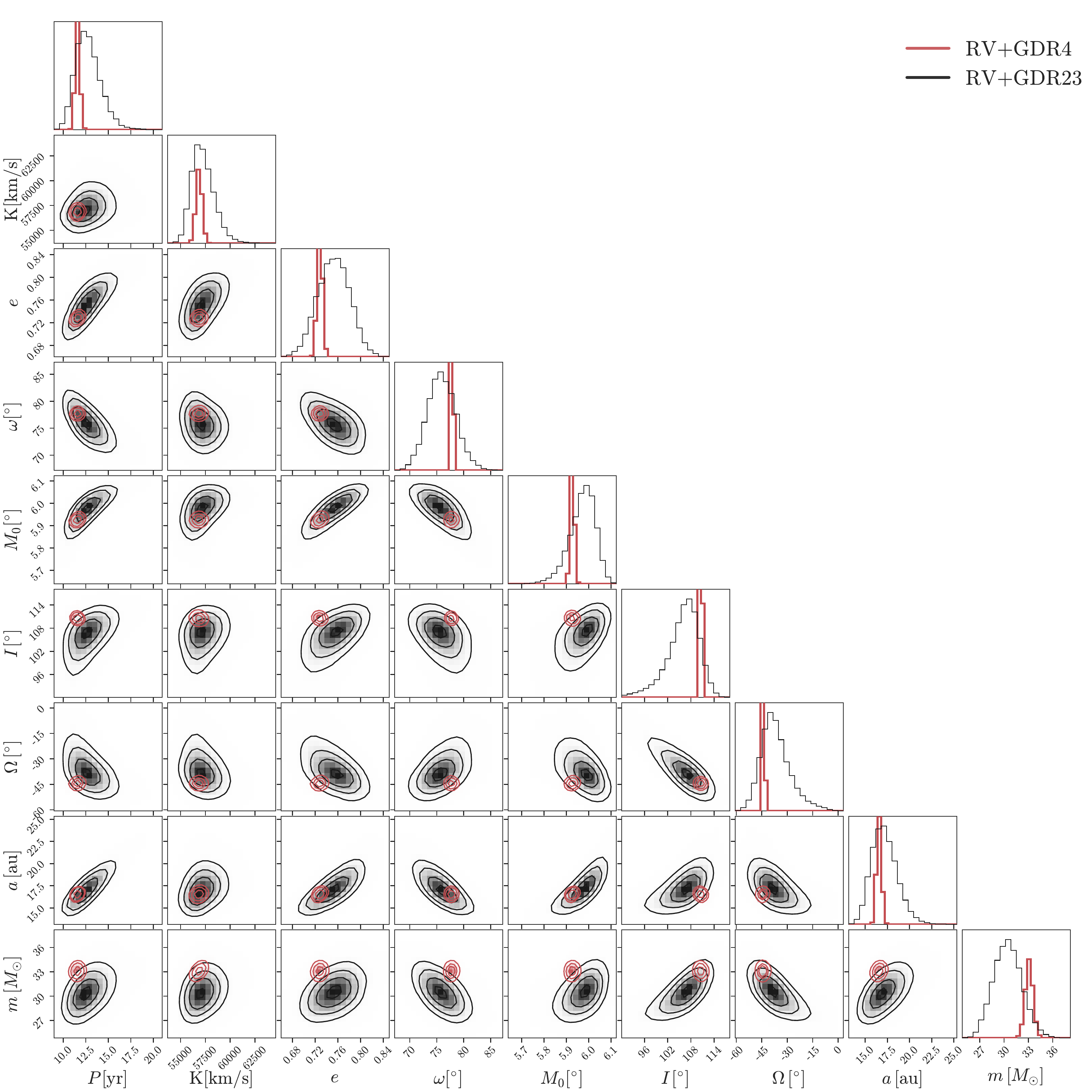}
    \caption{Comparison of the posterior distributions for Gaia BH3 system.}
    \label{fig:corner_bh3}
\end{figure*}

\bibliography{sample701}{}

\begin{thebibliography}{}
\expandafter\ifx\csname natexlab\endcsname\relax\def\natexlab#1{#1}\fi
\providecommand{\url}[1]{\href{#1}{#1}}
\providecommand{\dodoi}[1]{doi:~\href{http://doi.org/#1}{\nolinkurl{#1}}}
\providecommand{\doeprint}[1]{\href{http://ascl.net/#1}{\nolinkurl{http://ascl.net/#1}}}
\providecommand{\doarXiv}[1]{\href{https://arxiv.org/abs/#1}{\nolinkurl{https://arxiv.org/abs/#1}}}

\bibitem[{ESA(1997)}]{ESA1997}
 1997, ESA Special Publication, Vol. 1200, {The HIPPARCOS and TYCHO catalogues.
  Astrometric and photometric star catalogues derived from the ESA HIPPARCOS
  Space Astrometry Mission}

\bibitem[{J. {Andersen} {et~al.}(1983){Andersen}, {Clausen}, {Gimenez}, \&
  {Nordstroem}}]{Andersen1983A&A}
{Andersen}, J., {Clausen}, J.~V., {Gimenez}, A., \& {Nordstroem}, B. 1983,
  \bibinfo{title}{{Absolute dimensions of eclipsing binaries. II. The
  early-type semidetached system V Puppis.},} \aap, 128, 17

\bibitem[{J.~H. {Applegate}(1992){Applegate}}]{Applegate1992ApJ}
{Applegate}, J.~H. 1992, \bibinfo{title}{{A Mechanism for Orbital Period
  Modulation in Close Binaries},} \apj, 385, 621, \dodoi{10.1086/170967}

\bibitem[{ {Astropy Collaboration} {et~al.}(2013){Astropy Collaboration},
  {Robitaille}, {Tollerud}, {Greenfield}, {Droettboom}, {Bray}, {Aldcroft},
  {Davis}, {Ginsburg}, {Price-Whelan}, {Kerzendorf}, {Conley}, {Crighton},
  {Barbary}, {Muna}, {Ferguson}, {Grollier}, {Parikh}, {Nair}, {Unther},
  {Deil}, {Woillez}, {Conseil}, {Kramer}, {Turner}, {Singer}, {Fox}, {Weaver},
  {Zabalza}, {Edwards}, {Azalee Bostroem}, {Burke}, {Casey}, {Crawford},
  {Dencheva}, {Ely}, {Jenness}, {Labrie}, {Lim}, {Pierfederici}, {Pontzen},
  {Ptak}, {Refsdal}, {Servillat}, \& {Streicher}}]{2013A&A...558A..33A}
{Astropy Collaboration}, {Robitaille}, T.~P., {Tollerud}, E.~J., {et~al.} 2013,
  \bibinfo{title}{{Astropy: A community Python package for astronomy},} \aap,
  558, A33, \dodoi{10.1051/0004-6361/201322068}

\bibitem[{ {Astropy Collaboration} {et~al.}(2018){Astropy Collaboration},
  {Price-Whelan}, {Sip{\H{o}}cz}, {G{\"u}nther}, {Lim}, {Crawford}, {Conseil},
  {Shupe}, {Craig}, {Dencheva}, {Ginsburg}, {VanderPlas}, {Bradley},
  {P{\'e}rez-Su{\'a}rez}, {de Val-Borro}, {Aldcroft}, {Cruz}, {Robitaille},
  {Tollerud}, {Ardelean}, {Babej}, {Bach}, {Bachetti}, {Bakanov}, {Bamford},
  {Barentsen}, {Barmby}, {Baumbach}, {Berry}, {Biscani}, {Boquien}, {Bostroem},
  {Bouma}, {Brammer}, {Bray}, {Breytenbach}, {Buddelmeijer}, {Burke},
  {Calderone}, {Cano Rodr{\'\i}guez}, {Cara}, {Cardoso}, {Cheedella}, {Copin},
  {Corrales}, {Crichton}, {D'Avella}, {Deil}, {Depagne}, {Dietrich}, {Donath},
  {Droettboom}, {Earl}, {Erben}, {Fabbro}, {Ferreira}, {Finethy}, {Fox},
  {Garrison}, {Gibbons}, {Goldstein}, {Gommers}, {Greco}, {Greenfield},
  {Groener}, {Grollier}, {Hagen}, {Hirst}, {Homeier}, {Horton}, {Hosseinzadeh},
  {Hu}, {Hunkeler}, {Ivezi{\'c}}, {Jain}, {Jenness}, {Kanarek}, {Kendrew},
  {Kern}, {Kerzendorf}, {Khvalko}, {King}, {Kirkby}, {Kulkarni}, {Kumar},
  {Lee}, {Lenz}, {Littlefair}, {Ma}, {Macleod}, {Mastropietro}, {McCully},
  {Montagnac}, {Morris}, {Mueller}, {Mumford}, {Muna}, {Murphy}, {Nelson},
  {Nguyen}, {Ninan}, {N{\"o}the}, {Ogaz}, {Oh}, {Parejko}, {Parley}, {Pascual},
  {Patil}, {Patil}, {Plunkett}, {Prochaska}, {Rastogi}, {Reddy Janga},
  {Sabater}, {Sakurikar}, {Seifert}, {Sherbert}, {Sherwood-Taylor}, {Shih},
  {Sick}, {Silbiger}, {Singanamalla}, {Singer}, {Sladen}, {Sooley},
  {Sornarajah}, {Streicher}, {Teuben}, {Thomas}, {Tremblay}, {Turner},
  {Terr{\'o}n}, {van Kerkwijk}, {de la Vega}, {Watkins}, {Weaver}, {Whitmore},
  {Woillez}, {Zabalza}, \& {Astropy Contributors}}]{2018AJ....156..123A}
{Astropy Collaboration}, {Price-Whelan}, A.~M., {Sip{\H{o}}cz}, B.~M., {et~al.}
  2018, \bibinfo{title}{{The Astropy Project: Building an Open-science Project
  and Status of the v2.0 Core Package},} \aj, 156, 123,
  \dodoi{10.3847/1538-3881/aabc4f}

\bibitem[{ {Astropy Collaboration} {et~al.}(2022){Astropy Collaboration},
  {Price-Whelan}, {Lim}, {Earl}, {Starkman}, {Bradley}, {Shupe}, {Patil},
  {Corrales}, {Brasseur}, {N{\"o}the}, {Donath}, {Tollerud}, {Morris},
  {Ginsburg}, {Vaher}, {Weaver}, {Tocknell}, {Jamieson}, {van Kerkwijk},
  {Robitaille}, {Merry}, {Bachetti}, {G{\"u}nther}, {Aldcroft},
  {Alvarado-Montes}, {Archibald}, {B{\'o}di}, {Bapat}, {Barentsen},
  {Baz{\'a}n}, {Biswas}, {Boquien}, {Burke}, {Cara}, {Cara}, {Conroy},
  {Conseil}, {Craig}, {Cross}, {Cruz}, {D'Eugenio}, {Dencheva}, {Devillepoix},
  {Dietrich}, {Eigenbrot}, {Erben}, {Ferreira}, {Foreman-Mackey}, {Fox},
  {Freij}, {Garg}, {Geda}, {Glattly}, {Gondhalekar}, {Gordon}, {Grant},
  {Greenfield}, {Groener}, {Guest}, {Gurovich}, {Handberg}, {Hart},
  {Hatfield-Dodds}, {Homeier}, {Hosseinzadeh}, {Jenness}, {Jones}, {Joseph},
  {Kalmbach}, {Karamehmetoglu}, {Ka{\l}uszy{\'n}ski}, {Kelley}, {Kern},
  {Kerzendorf}, {Koch}, {Kulumani}, {Lee}, {Ly}, {Ma}, {MacBride}, {Maljaars},
  {Muna}, {Murphy}, {Norman}, {O'Steen}, {Oman}, {Pacifici}, {Pascual},
  {Pascual-Granado}, {Patil}, {Perren}, {Pickering}, {Rastogi}, {Roulston},
  {Ryan}, {Rykoff}, {Sabater}, {Sakurikar}, {Salgado}, {Sanghi}, {Saunders},
  {Savchenko}, {Schwardt}, {Seifert-Eckert}, {Shih}, {Jain}, {Shukla}, {Sick},
  {Simpson}, {Singanamalla}, {Singer}, {Singhal}, {Sinha}, {Sip{\H{o}}cz},
  {Spitler}, {Stansby}, {Streicher}, {{\v{S}}umak}, {Swinbank}, {Taranu},
  {Tewary}, {Tremblay}, {de Val-Borro}, {Van Kooten}, {Vasovi{\'c}}, {Verma},
  {de Miranda Cardoso}, {Williams}, {Wilson}, {Winkel}, {Wood-Vasey}, {Xue},
  {Yoachim}, {Zhang}, {Zonca}, \& {Astropy Project
  Contributors}}]{2022ApJ...935..167A}
{Astropy Collaboration}, {Price-Whelan}, A.~M., {Lim}, P.~L., {et~al.} 2022,
  \bibinfo{title}{{The Astropy Project: Sustaining and Growing a
  Community-oriented Open-source Project and the Latest Major Release (v5.0) of
  the Core Package},} \apj, 935, 167, \dodoi{10.3847/1538-4357/ac7c74}

\bibitem[{J.~N. {Bahcall} {et~al.}(1975){Bahcall}, {Charles}, {Davison},
  {Sanford}, {Kellogg}, \& {York}}]{Bahcall1975MNRAS}
{Bahcall}, J.~N., {Charles}, P.~A., {Davison}, P.~J.~N., {et~al.} 1975,
  \bibinfo{title}{{Copernicus - X-ray observations of 3U 9759-49.},} \mnras,
  171, 41P, \dodoi{10.1093/mnras/171.1.41P}

\bibitem[{S.~A. {Bell} {et~al.}(1987){Bell}, {Adamson}, \&
  {Hilditch}}]{Bell1987MNRAS}
{Bell}, S.~A., {Adamson}, A.~J., \& {Hilditch}, R.~W. 1987,
  \bibinfo{title}{{Simultaneous differential photometry with the ST Andrews
  twin photometric telescope - II. The eclipsing binaries SX Aurigae and TT
  Aurigae.},} \mnras, 224, 649, \dodoi{10.1093/mnras/224.3.649}

\bibitem[{L. {Binnendijk}(1960){Binnendijk}}]{Binnendijk1960}
{Binnendijk}, L. 1960, {Properties of double stars; a survey of parallaxes and
  orbits.}

\bibitem[{T. {Borkovits} {et~al.}(2011){Borkovits}, {Csizmadia},
  {Forg{\'a}cs-Dajka}, \& {Heged{\"u}s}}]{Borkovits2011A&A}
{Borkovits}, T., {Csizmadia}, S., {Forg{\'a}cs-Dajka}, E., \& {Heged{\"u}s}, T.
  2011, \bibinfo{title}{{Transit timing variations in eccentric hierarchical
  triple exoplanetary systems. I. Perturbations on the time scale of the
  orbital period of the perturber},} \aap, 528, A53,
  \dodoi{10.1051/0004-6361/201015867}

\bibitem[{T. {Borkovits} {et~al.}(2016){Borkovits}, {Hajdu}, {Sztakovics},
  {Rappaport}, {Levine}, {B{\'\i}r{\'o}}, \& {Klagyivik}}]{Borkovits2016MNRAS}
{Borkovits}, T., {Hajdu}, T., {Sztakovics}, J., {et~al.} 2016,
  \bibinfo{title}{{A comprehensive study of the Kepler triples via eclipse
  timing},} \mnras, 455, 4136, \dodoi{10.1093/mnras/stv2530}

\bibitem[{T. {Borkovits} {et~al.}(2015){Borkovits}, {Rappaport}, {Hajdu}, \&
  {Sztakovics}}]{Borkovits2015MNRAS}
{Borkovits}, T., {Rappaport}, S., {Hajdu}, T., \& {Sztakovics}, J. 2015,
  \bibinfo{title}{{Eclipse timing variation analyses of eccentric binaries with
  close tertiaries in the Kepler field},} \mnras, 448, 946,
  \dodoi{10.1093/mnras/stv015}

\bibitem[{T. {Borkovits} {et~al.}(2025){Borkovits}, {Rappaport}, {Mitnyan},
  {B{\'\i}r{\'o}}, {Cs{\'a}nyi}, {Forg{\'a}cs-Dajka}, {Forr{\'o}}, {Hajdu},
  {Seli}, {Sztakovics}, {G{\"o}bly{\"o}s}, \& {P{\'a}l}}]{Borkovits2025A&A}
{Borkovits}, T., {Rappaport}, S.~A., {Mitnyan}, T., {et~al.} 2025,
  \bibinfo{title}{{Then and now: A new look at the eclipse timing variations of
  hierarchical triple star candidates in the primordial Kepler field, revisited
  by TESS},} \aap, 695, A209, \dodoi{10.1051/0004-6361/202453616}

\bibitem[{W.~J. {Borucki} {et~al.}(2010){Borucki}, {Koch}, {Basri}, {Batalha},
  {Brown}, {Caldwell}, {Caldwell}, {Christensen-Dalsgaard}, {Cochran},
  {DeVore}, {Dunham}, {Dupree}, {Gautier}, {Geary}, {Gilliland}, {Gould},
  {Howell}, {Jenkins}, {Kondo}, {Latham}, {Marcy}, {Meibom}, {Kjeldsen},
  {Lissauer}, {Monet}, {Morrison}, {Sasselov}, {Tarter}, {Boss}, {Brownlee},
  {Owen}, {Buzasi}, {Charbonneau}, {Doyle}, {Fortney}, {Ford}, {Holman},
  {Seager}, {Steffen}, {Welsh}, {Rowe}, {Anderson}, {Buchhave}, {Ciardi},
  {Walkowicz}, {Sherry}, {Horch}, {Isaacson}, {Everett}, {Fischer}, {Torres},
  {Johnson}, {Endl}, {MacQueen}, {Bryson}, {Dotson}, {Haas}, {Kolodziejczak},
  {Van Cleve}, {Chandrasekaran}, {Twicken}, {Quintana}, {Clarke}, {Allen},
  {Li}, {Wu}, {Tenenbaum}, {Verner}, {Bruhweiler}, {Barnes}, \&
  {Prsa}}]{Borucki2010Sci}
{Borucki}, W.~J., {Koch}, D., {Basri}, G., {et~al.} 2010,
  \bibinfo{title}{{Kepler Planet-Detection Mission: Introduction and First
  Results},} Science, 327, 977, \dodoi{10.1126/science.1185402}

\bibitem[{T.~D. {Brandt}(2018){Brandt}}]{Brandt2018}
{Brandt}, T.~D. 2018, \bibinfo{title}{{The Hipparcos-Gaia Catalog of
  Accelerations},} \apjs, 239, 31, \dodoi{10.3847/1538-4365/aaec06}

\bibitem[{T.~D. {Brandt}(2021){Brandt}}]{Brandt2021}
{Brandt}, T.~D. 2021, \bibinfo{title}{{The Hipparcos-Gaia Catalog of
  Accelerations: Gaia EDR3 Edition},} \apjs, 254, 42,
  \dodoi{10.3847/1538-4365/abf93c}

\bibitem[{T.~D. {Brandt} {et~al.}(2021){Brandt}, {Dupuy}, {Li}, {Brandt},
  {Zeng}, {Michalik}, {Bardalez Gagliuffi}, \&
  {Raposo-Pulido}}]{Brandt2021_orvara}
{Brandt}, T.~D., {Dupuy}, T.~J., {Li}, Y., {et~al.} 2021,
  \bibinfo{title}{{orvara: An Efficient Code to Fit Orbits Using Radial
  Velocity, Absolute, and/or Relative Astrometry},} \aj, 162, 186,
  \dodoi{10.3847/1538-3881/ac042e}

\bibitem[{E. {Budding} {et~al.}(2021){Budding}, {Love}, {Blackford}, {Banks},
  \& {Rhodes}}]{Budding2021MNRAS}
{Budding}, E., {Love}, T., {Blackford}, M.~G., {Banks}, T., \& {Rhodes}, M.~J.
  2021, \bibinfo{title}{{Absolute parameters of young stars: V puppis},}
  \mnras, 502, 6032, \dodoi{10.1093/mnras/stab381}

\bibitem[{K.~E. {Conroy} {et~al.}(2014){Conroy}, {Pr{\v{s}}a}, {Stassun},
  {Orosz}, {Fabrycky}, \& {Welsh}}]{Conroy2014AJ}
{Conroy}, K.~E., {Pr{\v{s}}a}, A., {Stassun}, K.~G., {et~al.} 2014,
  \bibinfo{title}{{Kepler Eclipsing Binary Stars. IV. Precise Eclipse Times for
  Close Binaries and Identification of Candidate Three-body Systems},} \aj,
  147, 45, \dodoi{10.1088/0004-6256/147/2/45}

\bibitem[{F. {Feng} {et~al.}(2023){Feng}, {Butler}, {Vogt}, {Holden}, \&
  {Rui}}]{Feng2023MNRAS}
{Feng}, F., {Butler}, R.~P., {Vogt}, S.~S., {Holden}, B., \& {Rui}, Y. 2023,
  \bibinfo{title}{{Revised orbits of the two nearest Jupiters},} \mnras, 525,
  607, \dodoi{10.1093/mnras/stad2297}

\bibitem[{F. {Feng} {et~al.}(2019){Feng}, {Crane}, {Xuesong Wang}, {Teske},
  {Shectman}, {D{\'\i}az}, {Thompson}, {Jones}, \&
  {Butler}}]{Feng2019ApJS..242...25F}
{Feng}, F., {Crane}, J.~D., {Xuesong Wang}, S., {et~al.} 2019,
  \bibinfo{title}{{Search for Nearby Earth Analogs. I. 15 Planet Candidates
  Found in PFS Data},} \apjs, 242, 25, \dodoi{10.3847/1538-4365/ab1b16}

\bibitem[{F. {Feng} {et~al.}(2022){Feng}, {Butler}, {Vogt}, {Clement},
  {Tinney}, {Cui}, {Aizawa}, {Jones}, {Bailey}, {Burt}, {Carter}, {Crane},
  {Dotti}, {Holden}, {Ma}, {Ogihara}, {Oppenheimer}, {O'Toole}, {Shectman},
  {Wittenmyer}, {Wang}, {Wright}, \& {Xuan}}]{Feng2022}
{Feng}, F., {Butler}, R.~P., {Vogt}, S.~S., {et~al.} 2022, \bibinfo{title}{{3D
  Selection of 167 Substellar Companions to Nearby Stars},} \apjs, 262, 21,
  \dodoi{10.3847/1538-4365/ac7e57}

\bibitem[{ {Gaia Collaboration} {et~al.}(2018){Gaia Collaboration}, {Brown},
  {Vallenari}, {Prusti}, {de Bruijne}, {Babusiaux}, {Bailer-Jones}, {Biermann},
  {Evans}, {Eyer}, {Jansen}, {Jordi}, {Klioner}, {Lammers}, {Lindegren},
  {Luri}, {Mignard}, {Panem}, {Pourbaix}, {Randich}, {Sartoretti}, {Siddiqui},
  {Soubiran}, {van Leeuwen}, {Walton}, {Arenou}, {Bastian}, {Cropper},
  {Drimmel}, {Katz}, {Lattanzi}, {Bakker}, {Cacciari}, {Casta{\~n}eda},
  {Chaoul}, {Cheek}, {De Angeli}, {Fabricius}, {Guerra}, {Holl}, {Masana},
  {Messineo}, {Mowlavi}, {Nienartowicz}, {Panuzzo}, {Portell}, {Riello},
  {Seabroke}, {Tanga}, {Th{\'e}venin}, {Gracia-Abril}, {Comoretto},
  {Garcia-Reinaldos}, {Teyssier}, {Altmann}, {Andrae}, {Audard},
  {Bellas-Velidis}, {Benson}, {Berthier}, {Blomme}, {Burgess}, {Busso},
  {Carry}, {Cellino}, {Clementini}, {Clotet}, {Creevey}, {Davidson}, {De
  Ridder}, {Delchambre}, {Dell'Oro}, {Ducourant},
  {Fern{\'a}ndez-Hern{\'a}ndez}, {Fouesneau}, {Fr{\'e}mat}, {Galluccio},
  {Garc{\'\i}a-Torres}, {Gonz{\'a}lez-N{\'u}{\~n}ez}, {Gonz{\'a}lez-Vidal},
  {Gosset}, {Guy}, {Halbwachs}, {Hambly}, {Harrison}, {Hern{\'a}ndez},
  {Hestroffer}, {Hodgkin}, {Hutton}, {Jasniewicz}, {Jean-Antoine-Piccolo},
  {Jordan}, {Korn}, {Krone-Martins}, {Lanzafame}, {Lebzelter}, {L{\"o}ffler},
  {Manteiga}, {Marrese}, {Mart{\'\i}n-Fleitas}, {Moitinho}, {Mora}, {Muinonen},
  {Osinde}, {Pancino}, {Pauwels}, {Petit}, {Recio-Blanco}, {Richards},
  {Rimoldini}, {Robin}, {Sarro}, {Siopis}, {Smith}, {Sozzetti}, {S{\"u}veges},
  {Torra}, {van Reeven}, {Abbas}, {Abreu Aramburu}, {Accart}, {Aerts},
  {Altavilla}, {{\'A}lvarez}, {Alvarez}, {Alves}, {Anderson}, {Andrei},
  {Anglada Varela}, {Antiche}, {Antoja}, {Arcay}, {Astraatmadja}, {Bach},
  {Baker}, {Balaguer-N{\'u}{\~n}ez}, {Balm}, {Barache}, {Barata}, {Barbato},
  {Barblan}, {Barklem}, {Barrado}, {Barros}, {Barstow}, {Bartholom{\'e}
  Mu{\~n}oz}, {Bassilana}, {Becciani}, {Bellazzini}, {Berihuete}, {Bertone},
  {Bianchi}, {Bienaym{\'e}}, {Blanco-Cuaresma}, {Boch}, {Boeche}, {Bombrun},
  {Borrachero}, {Bossini}, {Bouquillon}, {Bourda}, {Bragaglia}, {Bramante},
  {Breddels}, {Bressan}, {Brouillet}, {Br{\"u}semeister}, {Brugaletta},
  {Bucciarelli}, {Burlacu}, {Busonero}, {Butkevich}, {Buzzi}, {Caffau},
  {Cancelliere}, {Cannizzaro}, {Cantat-Gaudin}, {Carballo}, {Carlucci},
  {Carrasco}, {Casamiquela}, {Castellani}, {Castro-Ginard}, {Charlot},
  {Chemin}, {Chiavassa}, {Cocozza}, {Costigan}, {Cowell}, {Crifo}, {Crosta},
  {Crowley}, {Cuypers}, {Dafonte}, {Damerdji}, {Dapergolas}, {David}, {David},
  {de Laverny}, {De Luise}, {De March}, {de Martino}, {de Souza}, {de Torres},
  {Debosscher}, {del Pozo}, {Delbo}, {Delgado}, {Delgado}, {Di Matteo},
  {Diakite}, {Diener}, {Distefano}, {Dolding}, {Drazinos}, {Dur{\'a}n},
  {Edvardsson}, {Enke}, {Eriksson}, {Esquej}, {Eynard Bontemps}, {Fabre},
  {Fabrizio}, {Faigler}, {Falc{\~a}o}, {Farr{\`a}s Casas}, {Federici},
  {Fedorets}, {Fernique}, {Figueras}, {Filippi}, {Findeisen}, {Fonti},
  {Fraile}, {Fraser}, {Fr{\'e}zouls}, {Gai}, {Galleti}, {Garabato},
  {Garc{\'\i}a-Sedano}, {Garofalo}, {Garralda}, {Gavel}, {Gavras}, {Gerssen},
  {Geyer}, {Giacobbe}, {Gilmore}, {Girona}, {Giuffrida}, {Glass}, {Gomes},
  {Granvik}, {Gueguen}, {Guerrier}, {Guiraud}, {Guti{\'e}rrez-S{\'a}nchez},
  {Haigron}, {Hatzidimitriou}, {Hauser}, {Haywood}, {Heiter}, {Helmi}, {Heu},
  {Hilger}, {Hobbs}, {Hofmann}, {Holland}, {Huckle}, {Hypki}, {Icardi},
  {Jan{\ss}en}, {Jevardat de Fombelle}, {Jonker}, {Juh{\'a}sz}, {Julbe},
  {Karampelas}, {Kewley}, {Klar}, {Kochoska}, {Kohley}, {Kolenberg},
  {Kontizas}, {Kontizas}, {Koposov}, {Kordopatis}, {Kostrzewa-Rutkowska},
  {Koubsky}, {Lambert}, {Lanza}, {Lasne}, {Lavigne}, {Le Fustec}, {Le
  Poncin-Lafitte}, {Lebreton}, {Leccia}, {Leclerc}, {Lecoeur-Taibi},
  {Lenhardt}, {Leroux}, {Liao}, {Licata}, {Lindstr{\o}m}, {Lister}, {Livanou},
  {Lobel}, {L{\'o}pez}, {Managau}, {Mann}, {Mantelet}, {Marchal}, {Marchant},
  {Marconi}, {Marinoni}, {Marschalk{\'o}}, {Marshall}, {Martino}, {Marton},
  {Mary}, {Massari}, {Matijevi{\v{c}}}, {Mazeh}, {McMillan}, {Messina},
  {Michalik}, {Millar}, {Molina}, {Molinaro}, {Moln{\'a}r}, {Montegriffo},
  {Mor}, {Morbidelli}, {Morel}, {Morris}, {Mulone}, {Muraveva}, {Musella},
  {Nelemans}, {Nicastro}, {Noval}, {O'Mullane}, {Ord{\'e}novic},
  {Ord{\'o}{\~n}ez-Blanco}, {Osborne}, {Pagani}, {Pagano}, {Pailler},
  {Palacin}, {Palaversa}, {Panahi}, {Pawlak}, {Piersimoni}, {Pineau}, {Plachy},
  {Plum}, {Poggio}, {Poujoulet}, {Pr{\v{s}}a}, {Pulone}, {Racero}, {Ragaini},
  {Rambaux}, {Ramos-Lerate}, {Regibo}, {Reyl{\'e}}, {Riclet}, {Ripepi}, {Riva},
  {Rivard}, {Rixon}, {Roegiers}, {Roelens}, {Romero-G{\'o}mez}, {Rowell},
  {Royer}, {Ruiz-Dern}, {Sadowski}, {Sagrist{\`a} Sell{\'e}s}, {Sahlmann},
  {Salgado}, {Salguero}, {Sanna}, {Santana-Ros}, {Sarasso}, {Savietto},
  {Schultheis}, {Sciacca}, {Segol}, {Segovia}, {S{\'e}gransan}, {Shih},
  {Siltala}, {Silva}, {Smart}, {Smith}, {Solano}, {Solitro}, {Sordo}, {Soria
  Nieto}, {Souchay}, {Spagna}, {Spoto}, {Stampa}, {Steele},
  {Steidelm{\"u}ller}, {Stephenson}, {Stoev}, {Suess}, {Surdej}, {Szabados},
  {Szegedi-Elek}, {Tapiador}, {Taris}, {Tauran}, {Taylor}, {Teixeira},
  {Terrett}, {Teyssandier}, {Thuillot}, {Titarenko}, {Torra Clotet}, {Turon},
  {Ulla}, {Utrilla}, {Uzzi}, {Vaillant}, {Valentini}, {Valette}, {van Elteren},
  {Van Hemelryck}, {van Leeuwen}, {Vaschetto}, {Vecchiato}, {Veljanoski},
  {Viala}, {Vicente}, {Vogt}, {von Essen}, {Voss}, {Votruba}, {Voutsinas},
  {Walmsley}, {Weiler}, {Wertz}, {Wevers}, {Wyrzykowski}, {Yoldas},
  {{\v{Z}}erjal}, {Ziaeepour}, {Zorec}, {Zschocke}, {Zucker}, {Zurbach}, \&
  {Zwitter}}]{GaiaCollaboration2018}
{Gaia Collaboration}, {Brown}, A.~G.~A., {Vallenari}, A., {et~al.} 2018,
  \bibinfo{title}{{Gaia Data Release 2. Summary of the contents and survey
  properties},} \aap, 616, A1, \dodoi{10.1051/0004-6361/201833051}

\bibitem[{ {Gaia Collaboration} {et~al.}(2023{\natexlab{a}}){Gaia
  Collaboration}, {Vallenari}, {Brown}, {Prusti}, {de Bruijne}, {Arenou},
  {Babusiaux}, {Biermann}, {Creevey}, {Ducourant}, {Evans}, {Eyer}, {Guerra},
  {Hutton}, {Jordi}, {Klioner}, {Lammers}, {Lindegren}, {Luri}, {Mignard},
  {Panem}, {Pourbaix}, {Randich}, {Sartoretti}, {Soubiran}, {Tanga}, {Walton},
  {Bailer-Jones}, {Bastian}, {Drimmel}, {Jansen}, {Katz}, {Lattanzi}, {van
  Leeuwen}, {Bakker}, {Cacciari}, {Casta{\~n}eda}, {De Angeli}, {Fabricius},
  {Fouesneau}, {Fr{\'e}mat}, {Galluccio}, {Guerrier}, {Heiter}, {Masana},
  {Messineo}, {Mowlavi}, {Nicolas}, {Nienartowicz}, {Pailler}, {Panuzzo},
  {Riclet}, {Roux}, {Seabroke}, {Sordo}, {Th{\'e}venin}, {Gracia-Abril},
  {Portell}, {Teyssier}, {Altmann}, {Andrae}, {Audard}, {Bellas-Velidis},
  {Benson}, {Berthier}, {Blomme}, {Burgess}, {Busonero}, {Busso},
  {C{\'a}novas}, {Carry}, {Cellino}, {Cheek}, {Clementini}, {Damerdji},
  {Davidson}, {de Teodoro}, {Nu{\~n}ez Campos}, {Delchambre}, {Dell'Oro},
  {Esquej}, {Fern{\'a}ndez-Hern{\'a}ndez}, {Fraile}, {Garabato},
  {Garc{\'\i}a-Lario}, {Gosset}, {Haigron}, {Halbwachs}, {Hambly}, {Harrison},
  {Hern{\'a}ndez}, {Hestroffer}, {Hodgkin}, {Holl}, {Jan{\ss}en}, {Jevardat de
  Fombelle}, {Jordan}, {Krone-Martins}, {Lanzafame}, {L{\"o}ffler}, {Marchal},
  {Marrese}, {Moitinho}, {Muinonen}, {Osborne}, {Pancino}, {Pauwels},
  {Recio-Blanco}, {Reyl{\'e}}, {Riello}, {Rimoldini}, {Roegiers}, {Rybizki},
  {Sarro}, {Siopis}, {Smith}, {Sozzetti}, {Utrilla}, {van Leeuwen}, {Abbas},
  {{\'A}brah{\'a}m}, {Abreu Aramburu}, {Aerts}, {Aguado}, {Ajaj},
  {Aldea-Montero}, {Altavilla}, {{\'A}lvarez}, {Alves}, {Anders}, {Anderson},
  {Anglada Varela}, {Antoja}, {Baines}, {Baker}, {Balaguer-N{\'u}{\~n}ez},
  {Balbinot}, {Balog}, {Barache}, {Barbato}, {Barros}, {Barstow},
  {Bartolom{\'e}}, {Bassilana}, {Bauchet}, {Becciani}, {Bellazzini},
  {Berihuete}, {Bernet}, {Bertone}, {Bianchi}, {Binnenfeld}, {Blanco-Cuaresma},
  {Blazere}, {Boch}, {Bombrun}, {Bossini}, {Bouquillon}, {Bragaglia},
  {Bramante}, {Breedt}, {Bressan}, {Brouillet}, {Brugaletta}, {Bucciarelli},
  {Burlacu}, {Butkevich}, {Buzzi}, {Caffau}, {Cancelliere}, {Cantat-Gaudin},
  {Carballo}, {Carlucci}, {Carnerero}, {Carrasco}, {Casamiquela}, {Castellani},
  {Castro-Ginard}, {Chaoul}, {Charlot}, {Chemin}, {Chiaramida}, {Chiavassa},
  {Chornay}, {Comoretto}, {Contursi}, {Cooper}, {Cornez}, {Cowell}, {Crifo},
  {Cropper}, {Crosta}, {Crowley}, {Dafonte}, {Dapergolas}, {David}, {David},
  {de Laverny}, {De Luise}, {De March}, {De Ridder}, {de Souza}, {de Torres},
  {del Peloso}, {del Pozo}, {Delbo}, {Delgado}, {Delisle}, {Demouchy},
  {Dharmawardena}, {Di Matteo}, {Diakite}, {Diener}, {Distefano}, {Dolding},
  {Edvardsson}, {Enke}, {Fabre}, {Fabrizio}, {Faigler}, {Fedorets}, {Fernique},
  {Fienga}, {Figueras}, {Fournier}, {Fouron}, {Fragkoudi}, {Gai},
  {Garcia-Gutierrez}, {Garcia-Reinaldos}, {Garc{\'\i}a-Torres}, {Garofalo},
  {Gavel}, {Gavras}, {Gerlach}, {Geyer}, {Giacobbe}, {Gilmore}, {Girona},
  {Giuffrida}, {Gomel}, {Gomez}, {Gonz{\'a}lez-N{\'u}{\~n}ez},
  {Gonz{\'a}lez-Santamar{\'\i}a}, {Gonz{\'a}lez-Vidal}, {Granvik}, {Guillout},
  {Guiraud}, {Guti{\'e}rrez-S{\'a}nchez}, {Guy}, {Hatzidimitriou}, {Hauser},
  {Haywood}, {Helmer}, {Helmi}, {Sarmiento}, {Hidalgo}, {Hilger},
  {H{\l}adczuk}, {Hobbs}, {Holland}, {Huckle}, {Jardine}, {Jasniewicz},
  {Jean-Antoine Piccolo}, {Jim{\'e}nez-Arranz}, {Jorissen}, {Juaristi
  Campillo}, {Julbe}, {Karbevska}, {Kervella}, {Khanna}, {Kontizas},
  {Kordopatis}, {Korn}, {K{\'o}sp{\'a}l}, {Kostrzewa-Rutkowska},
  {Kruszy{\'n}ska}, {Kun}, {Laizeau}, {Lambert}, {Lanza}, {Lasne}, {Le
  Campion}, {Lebreton}, {Lebzelter}, {Leccia}, {Leclerc}, {Lecoeur-Taibi},
  {Liao}, {Licata}, {Lindstr{\o}m}, {Lister}, {Livanou}, {Lobel}, {Lorca},
  {Loup}, {Madrero Pardo}, {Magdaleno Romeo}, {Managau}, {Mann}, {Manteiga},
  {Marchant}, {Marconi}, {Marcos}, {Marcos Santos}, {Mar{\'\i}n Pina},
  {Marinoni}, {Marocco}, {Marshall}, {Martin Polo}, {Mart{\'\i}n-Fleitas},
  {Marton}, {Mary}, {Masip}, {Massari}, {Mastrobuono-Battisti}, {Mazeh},
  {McMillan}, {Messina}, {Michalik}, {Millar}, {Mints}, {Molina}, {Molinaro},
  {Moln{\'a}r}, {Monari}, {Mongui{\'o}}, {Montegriffo}, {Montero}, {Mor},
  {Mora}, {Morbidelli}, {Morel}, {Morris}, {Muraveva}, {Murphy}, {Musella},
  {Nagy}, {Noval}, {Oca{\~n}a}, {Ogden}, {Ordenovic}, {Osinde}, {Pagani},
  {Pagano}, {Palaversa}, {Palicio}, {Pallas-Quintela}, {Panahi},
  {Payne-Wardenaar}, {Pe{\~n}alosa Esteller}, {Penttil{\"a}}, {Pichon},
  {Piersimoni}, {Pineau}, {Plachy}, {Plum}, {Poggio}, {Pr{\v{s}}a}, {Pulone},
  {Racero}, {Ragaini}, {Rainer}, {Raiteri}, {Rambaux}, {Ramos}, {Ramos-Lerate},
  {Re Fiorentin}, {Regibo}, {Richards}, {Rios Diaz}, {Ripepi}, {Riva}, {Rix},
  {Rixon}, {Robichon}, {Robin}, {Robin}, {Roelens}, {Rogues}, {Rohrbasser},
  {Romero-G{\'o}mez}, {Rowell}, {Royer}, {Ruz Mieres}, {Rybicki}, {Sadowski},
  {S{\'a}ez N{\'u}{\~n}ez}, {Sagrist{\`a} Sell{\'e}s}, {Sahlmann}, {Salguero},
  {Samaras}, {Sanchez Gimenez}, {Sanna}, {Santove{\~n}a}, {Sarasso},
  {Schultheis}, {Sciacca}, {Segol}, {Segovia}, {S{\'e}gransan}, {Semeux},
  {Shahaf}, {Siddiqui}, {Siebert}, {Siltala}, {Silvelo}, {Slezak}, {Slezak},
  {Smart}, {Snaith}, {Solano}, {Solitro}, {Souami}, {Souchay}, {Spagna},
  {Spina}, {Spoto}, {Steele}, {Steidelm{\"u}ller}, {Stephenson}, {S{\"u}veges},
  {Surdej}, {Szabados}, {Szegedi-Elek}, {Taris}, {Taylor}, {Teixeira},
  {Tolomei}, {Tonello}, {Torra}, {Torra}, {Torralba Elipe}, {Trabucchi},
  {Tsounis}, {Turon}, {Ulla}, {Unger}, {Vaillant}, {van Dillen}, {van Reeven},
  {Vanel}, {Vecchiato}, {Viala}, {Vicente}, {Voutsinas}, {Weiler}, {Wevers},
  {Wyrzykowski}, {Yoldas}, {Yvard}, {Zhao}, {Zorec}, {Zucker}, \&
  {Zwitter}}]{GaiaCollaboration2023}
{Gaia Collaboration}, {Vallenari}, A., {Brown}, A.~G.~A., {et~al.}
  2023{\natexlab{a}}, \bibinfo{title}{{Gaia Data Release 3. Summary of the
  content and survey properties},} \aap, 674, A1,
  \dodoi{10.1051/0004-6361/202243940}

\bibitem[{ {Gaia Collaboration} {et~al.}(2023{\natexlab{b}}){Gaia
  Collaboration}, {Arenou}, {Babusiaux}, {Barstow}, {Faigler}, {Jorissen},
  {Kervella}, {Mazeh}, {Mowlavi}, {Panuzzo}, {Sahlmann}, {Shahaf}, {Sozzetti},
  {Bauchet}, {Damerdji}, {Gavras}, {Giacobbe}, {Gosset}, {Halbwachs}, {Holl},
  {Lattanzi}, {Leclerc}, {Morel}, {Pourbaix}, {Re Fiorentin}, {Sadowski},
  {S{\'e}gransan}, {Siopis}, {Teyssier}, {Zwitter}, {Planquart}, {Brown},
  {Vallenari}, {Prusti}, {de Bruijne}, {Biermann}, {Creevey}, {Ducourant},
  {Evans}, {Eyer}, {Guerra}, {Hutton}, {Jordi}, {Klioner}, {Lammers},
  {Lindegren}, {Luri}, {Mignard}, {Panem}, {Randich}, {Sartoretti}, {Soubiran},
  {Tanga}, {Walton}, {Bailer-Jones}, {Bastian}, {Drimmel}, {Jansen}, {Katz},
  {van Leeuwen}, {Bakker}, {Cacciari}, {Casta{\~n}eda}, {De Angeli},
  {Fabricius}, {Fouesneau}, {Fr{\'e}mat}, {Galluccio}, {Guerrier}, {Heiter},
  {Masana}, {Messineo}, {Nicolas}, {Nienartowicz}, {Pailler}, {Riclet}, {Roux},
  {Seabroke}, {Sordo}, {Th{\'e}venin}, {Gracia-Abril}, {Portell}, {Altmann},
  {Andrae}, {Audard}, {Bellas-Velidis}, {Benson}, {Berthier}, {Blomme},
  {Burgess}, {Busonero}, {Busso}, {C{\'a}novas}, {Carry}, {Cellino}, {Cheek},
  {Clementini}, {Davidson}, {de Teodoro}, {Nu{\~n}ez Campos}, {Delchambre},
  {Dell'Oro}, {Esquej}, {Fern{\'a}ndez-Hern{\'a}ndez}, {Fraile}, {Garabato},
  {Garc{\'\i}a-Lario}, {Haigron}, {Hambly}, {Harrison}, {Hern{\'a}ndez},
  {Hestroffer}, {Hodgkin}, {Jan{\ss}en}, {Jevardat de Fombelle}, {Jordan},
  {Krone-Martins}, {Lanzafame}, {L{\"o}ffler}, {Marchal}, {Marrese},
  {Moitinho}, {Muinonen}, {Osborne}, {Pancino}, {Pauwels}, {Recio-Blanco},
  {Reyl{\'e}}, {Riello}, {Rimoldini}, {Roegiers}, {Rybizki}, {Sarro}, {Smith},
  {Utrilla}, {van Leeuwen}, {Abbas}, {{\'A}brah{\'a}m}, {Abreu Aramburu},
  {Aerts}, {Aguado}, {Ajaj}, {Aldea-Montero}, {Altavilla}, {{\'A}lvarez},
  {Alves}, {Anders}, {Anderson}, {Anglada Varela}, {Antoja}, {Baines}, {Baker},
  {Balaguer-N{\'u}{\~n}ez}, {Balbinot}, {Balog}, {Barache}, {Barbato},
  {Barros}, {Bartolom{\'e}}, {Bassilana}, {Becciani}, {Bellazzini},
  {Berihuete}, {Bernet}, {Bertone}, {Bianchi}, {Binnenfeld}, {Blanco-Cuaresma},
  {Blazere}, {Boch}, {Bombrun}, {Bossini}, {Bouquillon}, {Bragaglia},
  {Bramante}, {Breedt}, {Bressan}, {Brouillet}, {Brugaletta}, {Bucciarelli},
  {Burlacu}, {Butkevich}, {Buzzi}, {Caffau}, {Cancelliere}, {Cantat-Gaudin},
  {Carballo}, {Carlucci}, {Carnerero}, {Carrasco}, {Casamiquela}, {Castellani},
  {Castro-Ginard}, {Chaoul}, {Charlot}, {Chemin}, {Chiaramida}, {Chiavassa},
  {Chornay}, \& {Comoretto}}]{GaiaCollaboration2023_ruweA&A}
{Gaia Collaboration}, {Arenou}, F., {Babusiaux}, C., {et~al.}
  2023{\natexlab{b}}, \bibinfo{title}{{Gaia Data Release 3. Stellar
  multiplicity, a teaser for the hidden treasure},} \aap, 674, A34,
  \dodoi{10.1051/0004-6361/202243782}

\bibitem[{ {Gaia Collaboration} {et~al.}(2024){Gaia Collaboration}, {Panuzzo},
  {Mazeh}, {Arenou}, {Holl}, {Caffau}, {Jorissen}, {Babusiaux}, {Gavras},
  {Sahlmann}, {Bastian}, {Wyrzykowski}, {Eyer}, {Leclerc}, {Bauchet},
  {Bombrun}, {Mowlavi}, {Seabroke}, {Teyssier}, {Balbinot}, {Helmi}, {Brown},
  {Vallenari}, {Prusti}, {de Bruijne}, {Barbier}, {Biermann}, {Creevey},
  {Ducourant}, {Evans}, {Guerra}, {Hutton}, {Jordi}, {Klioner}, {Lammers},
  {Lindegren}, {Luri}, {Mignard}, {Nicolas}, {Randich}, {Sartoretti},
  {Smiljanic}, {Tanga}, {Walton}, {Aerts}, {Bailer-Jones}, {Cropper},
  {Drimmel}, {Jansen}, {Katz}, {Lattanzi}, {Soubiran}, {Th{\'e}venin}, {van
  Leeuwen}, {Andrae}, {Audard}, {Bakker}, {Blomme}, {Casta{\~n}eda}, {De
  Angeli}, {Fabricius}, {Fouesneau}, {Fr{\'e}mat}, {Galluccio}, {Guerrier},
  {Heiter}, {Masana}, {Messineo}, {Nienartowicz}, {Pailler}, {Riclet}, {Roux},
  {Sordo}, {Gracia-Abril}, {Portell}, {Altmann}, {Benson}, {Berthier},
  {Burgess}, {Busonero}, {Busso}, {Cacciari}, {C{\'a}novas}, {Carrasco},
  {Carry}, {Cellino}, {Cheek}, {Clementini}, {Damerdji}, {Davidson}, {de
  Teodoro}, {Delchambre}, {Dell'Oro}, {Fraile Garcia}, {Garabato},
  {Garc{\'\i}a-Lario}, {Haigron}, {Hambly}, {Harrison}, {Hatzidimitriou},
  {Hern{\'a}ndez}, {Hestroffer}, {Hodgkin}, {Jamal}, {Jevardat de Fombelle},
  {Jordan}, {Krone-Martins}, {Lanzafame}, {L{\"o}ffler}, {Lorca}, {Marchal},
  {Marrese}, {Moitinho}, {Muinonen}, {Nu{\~n}ez Campos}, {Oreshina-Slezak},
  {Osborne}, {Pancino}, {Pauwels}, {Recio-Blanco}, {Riello}, {Rimoldini},
  {Robin}, {Roegiers}, {Sarro}, {Schultheis}, {Smith}, {Sozzetti}, {Utrilla},
  {van Leeuwen}, {Weingrill}, {Abbas}, {{\'A}brah{\'a}m}, {Abreu Aramburu},
  {Ahmed}, {Altavilla}, {{\'A}lvarez}, {Anders}, {Anderson}, {Anglada Varela},
  {Antoja}, {Baig}, {Baines}, {Baker}, {Balaguer-N{\'u}{\~n}ez}, {Balog},
  {Barache}, {Barros}, {Barstow}, {Bartolom{\'e}}, {Bashi}, {Bassilana},
  {Baudeau}, {Becciani}, {Bedin}, {Bellas-Velidis}, {Bellazzini}, {Beordo},
  {Bernet}, {Bertolotto}, {Bertone}, {Bianchi}, {Binnenfeld},
  {Blanco-Cuaresma}, {Bland-Hawthorn}, {Blazere}, {Boch}, {Bossini},
  {Bouquillon}, {Bragaglia}, {Braine}, {Bratsolis}, {Breedt}, {Bressan},
  {Brouillet}, {Brugaletta}, {Bucciarelli}, {Butkevich}, {Buzzi}, {Camut},
  {Cancelliere}, {Cantat-Gaudin}, {Capilla Guilarte}, {Carballo}, {Carlucci},
  {Carnerero}, {Carretero}, {Carton}, {Casamiquela}, {Casey}, {Castellani},
  {Castro-Ginard}, {Ceraj}, {Cesare}, {Charlot}, {Chaudet}, {Chemin},
  {Chiavassa}, {Chornay}, \& {Chosson}}]{GaiaBH3_2024}
{Gaia Collaboration}, {Panuzzo}, P., {Mazeh}, T., {et~al.} 2024,
  \bibinfo{title}{{Discovery of a dormant 33 solar-mass black hole in
  pre-release Gaia astrometry},} \aap, 686, L2,
  \dodoi{10.1051/0004-6361/202449763}

\bibitem[{R. {Giacconi} {et~al.}(1974){Giacconi}, {Murray}, {Gursky},
  {Kellogg}, {Schreier}, {Matilsky}, {Koch}, \& {Tananbaum}}]{Giacconi1974ApJS}
{Giacconi}, R., {Murray}, S., {Gursky}, H., {et~al.} 1974, \bibinfo{title}{{The
  Third UHURU Catalog of X-Ray Sources},} \apjs, 27, 37, \dodoi{10.1086/190288}

\bibitem[{D.~R. {Gies} {et~al.}(2015){Gies}, {Matson}, {Guo}, {Lester},
  {Orosz}, \& {Peters}}]{Gies2015AJ}
{Gies}, D.~R., {Matson}, R.~A., {Guo}, Z., {et~al.} 2015,
  \bibinfo{title}{{Kepler Eclipsing Binaries with Stellar Companions},} \aj,
  150, 178, \dodoi{10.1088/0004-6256/150/6/178}

\bibitem[{W.~D. {Heintz}(1978){Heintz}}]{Heintz1978}
{Heintz}, W.~D. 1978, {Double stars}, Vol.~15

\bibitem[{B. {Holl} {et~al.}(2023){Holl}, {Sozzetti}, {Sahlmann}, {Giacobbe},
  {S{\'e}gransan}, {Unger}, {Delisle}, {Barbato}, {Lattanzi}, {Morbidelli}, \&
  {Sosnowska}}]{Holl2023A&A}
{Holl}, B., {Sozzetti}, A., {Sahlmann}, J., {et~al.} 2023,
  \bibinfo{title}{{Gaia Data Release 3. Astrometric orbit determination with
  Markov chain Monte Carlo and genetic algorithms: Systems with stellar,
  sub-stellar, and planetary mass companions},} \aap, 674, A10,
  \dodoi{10.1051/0004-6361/202244161}

\bibitem[{J.~D. Hunter(2007)Hunter}]{Hunter:2007}
Hunter, J.~D. 2007, \bibinfo{title}{Matplotlib: A 2D graphics environment,}
  Computing in Science \& Engineering, 9, 90, \dodoi{10.1109/MCSE.2007.55}

\bibitem[{J.~R.~P. {In{\'a}cio} {et~al.}(2024){In{\'a}cio}, {Mac{\^e}do},
  {Ferreira}, {Lisboa}, {Mendes}, {Pereira}, {da Silva}, \&
  {Almeida}}]{Inacio2024MNRAS}
{In{\'a}cio}, J. R.~P., {Mac{\^e}do}, I.~M., {Ferreira}, {\'E}. V.~X., {et~al.}
  2024, \bibinfo{title}{{Search and characterization of third-body candidates
  around short-period binaries using Kepler and TESS data},} \mnras, 529, 2967,
  \dodoi{10.1093/mnras/stae357}

\bibitem[{P. {Kervella} {et~al.}(2019){Kervella}, {Arenou}, {Mignard}, \&
  {Th{\'e}venin}}]{Kervella2019}
{Kervella}, P., {Arenou}, F., {Mignard}, F., \& {Th{\'e}venin}, F. 2019,
  \bibinfo{title}{{Stellar and substellar companions of nearby stars from Gaia
  DR2. Binarity from proper motion anomaly},} \aap, 623, A72,
  \dodoi{10.1051/0004-6361/201834371}

\bibitem[{P. {Kervella} {et~al.}(2022){Kervella}, {Arenou}, \&
  {Th{\'e}venin}}]{Kervella2022}
{Kervella}, P., {Arenou}, F., \& {Th{\'e}venin}, F. 2022,
  \bibinfo{title}{{Stellar and substellar companions from Gaia EDR3.
  Proper-motion anomaly and resolved common proper-motion pairs},} \aap, 657,
  A7, \dodoi{10.1051/0004-6361/202142146}

\bibitem[{F. {Kiefer} {et~al.}(2024){Kiefer}, {Lagrange}, {Rubini}, \&
  {Philipot}}]{Kiefer2024arXiv}
{Kiefer}, F., {Lagrange}, A.-M., {Rubini}, P., \& {Philipot}, F. 2024,
  \bibinfo{title}{{Searching for substellar companion candidates with Gaia. I.
  Introducing the GaiaPMEX tool},} arXiv e-prints, arXiv:2409.16992,
  \dodoi{10.48550/arXiv.2409.16992}

\bibitem[{R.~H. {Koch} {et~al.}(1981){Koch}, {Bradstreet}, {Perry}, \&
  {Pfeiffer}}]{Koch1981PASP}
{Koch}, R.~H., {Bradstreet}, D.~H., {Perry}, P.~M., \& {Pfeiffer}, R.~J. 1981,
  \bibinfo{title}{{IUE spectra of the hot close binary V Pup.},} \pasp, 93,
  621, \dodoi{10.1086/130897}

\bibitem[{K. {Li} {et~al.}(2021){Li}, {Xia}, {Kim}, {Hu}, {Guo}, {Jeong},
  {Chen}, \& {Gao}}]{LiKai2021ApJ}
{Li}, K., {Xia}, Q.-Q., {Kim}, C.-H., {et~al.} 2021, \bibinfo{title}{{Two
  Contact Binaries with Mass Ratios Close to the Minimum Mass Ratio},} \apj,
  922, 122, \dodoi{10.3847/1538-4357/ac242f}

\bibitem[{L. {Lindegren} {et~al.}(2012){Lindegren}, {Lammers}, {Hobbs},
  {O'Mullane}, {Bastian}, \& {Hern{\'a}ndez}}]{Lindegren2012A&A}
{Lindegren}, L., {Lammers}, U., {Hobbs}, D., {et~al.} 2012,
  \bibinfo{title}{{The astrometric core solution for the Gaia mission. Overview
  of models, algorithms, and software implementation},} \aap, 538, A78,
  \dodoi{10.1051/0004-6361/201117905}

\bibitem[{L. {Lindegren} {et~al.}(2018){Lindegren}, {Hern{\'a}ndez}, {Bombrun},
  {Klioner}, {Bastian}, {Ramos-Lerate}, {de Torres}, {Steidelm{\"u}ller},
  {Stephenson}, {Hobbs}, {Lammers}, {Biermann}, {Geyer}, {Hilger}, {Michalik},
  {Stampa}, {McMillan}, {Casta{\~n}eda}, {Clotet}, {Comoretto}, {Davidson},
  {Fabricius}, {Gracia}, {Hambly}, {Hutton}, {Mora}, {Portell}, {van Leeuwen},
  {Abbas}, {Abreu}, {Altmann}, {Andrei}, {Anglada}, {Balaguer-N{\'u}{\~n}ez},
  {Barache}, {Becciani}, {Bertone}, {Bianchi}, {Bouquillon}, {Bourda},
  {Br{\"u}semeister}, {Bucciarelli}, {Busonero}, {Buzzi}, {Cancelliere},
  {Carlucci}, {Charlot}, {Cheek}, {Crosta}, {Crowley}, {de Bruijne}, {de
  Felice}, {Drimmel}, {Esquej}, {Fienga}, {Fraile}, {Gai}, {Garralda},
  {Gonz{\'a}lez-Vidal}, {Guerra}, {Hauser}, {Hofmann}, {Holl}, {Jordan},
  {Lattanzi}, {Lenhardt}, {Liao}, {Licata}, {Lister}, {L{\"o}ffler},
  {Marchant}, {Martin-Fleitas}, {Messineo}, {Mignard}, {Morbidelli}, {Poggio},
  {Riva}, {Rowell}, {Salguero}, {Sarasso}, {Sciacca}, {Siddiqui}, {Smart},
  {Spagna}, {Steele}, {Taris}, {Torra}, {van Elteren}, {van Reeven}, \&
  {Vecchiato}}]{Lindegren2018}
{Lindegren}, L., {Hern{\'a}ndez}, J., {Bombrun}, A., {et~al.} 2018,
  \bibinfo{title}{{Gaia Data Release 2. The astrometric solution},} \aap, 616,
  A2, \dodoi{10.1051/0004-6361/201832727}

\bibitem[{T.~J. {Maccarone} {et~al.}(2009){Maccarone}, {Fender}, {Knigge}, \&
  {Tzioumis}}]{Maccarone2009MNRAS}
{Maccarone}, T.~J., {Fender}, R.~P., {Knigge}, C., \& {Tzioumis}, A.~K. 2009,
  \bibinfo{title}{{Constraints on black hole accretion in V Puppis},} \mnras,
  393, 1070, \dodoi{10.1111/j.1365-2966.2008.14291.x}

\bibitem[{T. {Mitnyan} {et~al.}(2024){Mitnyan}, {Borkovits}, {Czavalinga},
  {Rappaport}, {P{\'a}l}, {Powell}, \& {Hajdu}}]{Mitnyan2024A&A}
{Mitnyan}, T., {Borkovits}, T., {Czavalinga}, D.~R., {et~al.} 2024,
  \bibinfo{title}{{Eclipse-timing study of new hierarchical triple star
  candidates in the northern continuous viewing zone of TESS},} \aap, 685, A43,
  \dodoi{10.1051/0004-6361/202348909}

\bibitem[{N. {Nanouris} {et~al.}(2015){Nanouris}, {Kalimeris}, {Antonopoulou},
  \& {Rovithis-Livaniou}}]{Nanouris2015A&A}
{Nanouris}, N., {Kalimeris}, A., {Antonopoulou}, E., \& {Rovithis-Livaniou}, H.
  2015, \bibinfo{title}{{Efficiency of ETV diagrams as diagnostic tools for
  long-term period variations. II. Non-conservative mass transfer, and
  gravitational radiation},} \aap, 575, A64,
  \dodoi{10.1051/0004-6361/201323136}

\bibitem[{M.~J. {Pecaut} \& E.~E. {Mamajek}(2013){Pecaut} \&
  {Mamajek}}]{mamajek2013ApJS}
{Pecaut}, M.~J., \& {Mamajek}, E.~E. 2013, \bibinfo{title}{{Intrinsic Colors,
  Temperatures, and Bolometric Corrections of Pre-main-sequence Stars},} \apjs,
  208, 9, \dodoi{10.1088/0067-0049/208/1/9}

\bibitem[{A. {Popowicz} {et~al.}(2017){Popowicz}, {Pigulski}, {Bernacki},
  {Kuschnig}, {Pablo}, {Ramiaramanantsoa}, {Zoc{\l}o{\'n}ska}, {Baade},
  {Handler}, {Moffat}, {Wade}, {Neiner}, {Rucinski}, {Weiss}, {Koudelka},
  {Orlea{\'n}ski}, {Schwarzenberg-Czerny}, \& {Zwintz}}]{Popowicz2017A&A_BRITE}
{Popowicz}, A., {Pigulski}, A., {Bernacki}, K., {et~al.} 2017,
  \bibinfo{title}{{BRITE Constellation: data processing and photometry},} \aap,
  605, A26, \dodoi{10.1051/0004-6361/201730806}

\bibitem[{S. {Qian}(2001){Qian}}]{Qian2001MNRAS}
{Qian}, S. 2001, \bibinfo{title}{{Orbital period changes of contact binary
  systems: direct evidence for thermal relaxation oscillation theory},} \mnras,
  328, 914, \dodoi{10.1046/j.1365-8711.2001.04921.x}

\bibitem[{S.~B. {Qian} {et~al.}(2008){Qian}, {Liao}, \& {Fern{\'a}ndez
  Laj{\'u}s}}]{Qian2008ApJ}
{Qian}, S.~B., {Liao}, W.~P., \& {Fern{\'a}ndez Laj{\'u}s}, E. 2008,
  \bibinfo{title}{{Evidence of a Massive Black Hole Companion in the Massive
  Eclipsing Binary V Puppis},} \apj, 687, 466, \dodoi{10.1086/591515}

\bibitem[{S. {Rappaport} {et~al.}(2013){Rappaport}, {Deck}, {Levine},
  {Borkovits}, {Carter}, {El Mellah}, {Sanchis-Ojeda}, \&
  {Kalomeni}}]{Rappaport2013ApJ}
{Rappaport}, S., {Deck}, K., {Levine}, A., {et~al.} 2013,
  \bibinfo{title}{{Triple-star Candidates among the Kepler Binaries},} \apj,
  768, 33, \dodoi{10.1088/0004-637X/768/1/33}

\bibitem[{G.~R. {Ricker} {et~al.}(2015){Ricker}, {Winn}, {Vanderspek},
  {Latham}, {Bakos}, {Bean}, {Berta-Thompson}, {Brown}, {Buchhave}, {Butler},
  {Butler}, {Chaplin}, {Charbonneau}, {Christensen-Dalsgaard}, {Clampin},
  {Deming}, {Doty}, {De Lee}, {Dressing}, {Dunham}, {Endl}, {Fressin}, {Ge},
  {Henning}, {Holman}, {Howard}, {Ida}, {Jenkins}, {Jernigan}, {Johnson},
  {Kaltenegger}, {Kawai}, {Kjeldsen}, {Laughlin}, {Levine}, {Lin}, {Lissauer},
  {MacQueen}, {Marcy}, {McCullough}, {Morton}, {Narita}, {Paegert}, {Palle},
  {Pepe}, {Pepper}, {Quirrenbach}, {Rinehart}, {Sasselov}, {Sato}, {Seager},
  {Sozzetti}, {Stassun}, {Sullivan}, {Szentgyorgyi}, {Torres}, {Udry}, \&
  {Villasenor}}]{Ricker2015}
{Ricker}, G.~R., {Winn}, J.~N., {Vanderspek}, R., {et~al.} 2015,
  \bibinfo{title}{{Transiting Exoplanet Survey Satellite (TESS)},} Journal of
  Astronomical Telescopes, Instruments, and Systems, 1, 014003,
  \dodoi{10.1117/1.JATIS.1.1.014003}

\bibitem[{D.~P. {Schneider} {et~al.}(1979){Schneider}, {Darland}, \&
  {Leung}}]{Schneider1979AJ}
{Schneider}, D.~P., {Darland}, J.~J., \& {Leung}, K.~C. 1979,
  \bibinfo{title}{{Semidetached systems of spectral type B: BF Aurigae,
  {\textmu}$^{1}$ Scorpii, and V Puppis.},} \aj, 84, 236,
  \dodoi{10.1086/112412}

\bibitem[{J.~C. {Smith} {et~al.}(2012){Smith}, {Stumpe}, {Van Cleve},
  {Jenkins}, {Barclay}, {Fanelli}, {Girouard}, {Kolodziejczak}, {McCauliff},
  {Morris}, \& {Twicken}}]{Smith2012PASP}
{Smith}, J.~C., {Stumpe}, M.~C., {Van Cleve}, J.~E., {et~al.} 2012,
  \bibinfo{title}{{Kepler Presearch Data Conditioning II - A Bayesian Approach
  to Systematic Error Correction},} \pasp, 124, 1000, \dodoi{10.1086/667697}

\bibitem[{M.~C. {Stumpe} {et~al.}(2014){Stumpe}, {Smith}, {Catanzarite}, {Van
  Cleve}, {Jenkins}, {Twicken}, \& {Girouard}}]{Stumpe2014PASP}
{Stumpe}, M.~C., {Smith}, J.~C., {Catanzarite}, J.~H., {et~al.} 2014,
  \bibinfo{title}{{Multiscale Systematic Error Correction via Wavelet-Based
  Bandsplitting in Kepler Data},} \pasp, 126, 100, \dodoi{10.1086/674989}

\bibitem[{M.~C. {Stumpe} {et~al.}(2012){Stumpe}, {Smith}, {Van Cleve},
  {Twicken}, {Barclay}, {Fanelli}, {Girouard}, {Jenkins}, {Kolodziejczak},
  {McCauliff}, \& {Morris}}]{Stumpe2012PASP}
{Stumpe}, M.~C., {Smith}, J.~C., {Van Cleve}, J.~E., {et~al.} 2012,
  \bibinfo{title}{{Kepler Presearch Data Conditioning
  I{\textemdash}Architecture and Algorithms for Error Correction in Kepler
  Light Curves},} \pasp, 124, 985, \dodoi{10.1086/667698}

\bibitem[{T.~N. {Thiele}(1883){Thiele}}]{Thiele1883}
{Thiele}, T.~N. 1883, \bibinfo{title}{{Neue Methode zur Berechung von
  Doppelsternbahnen},} Astronomische Nachrichten, 104, 245

\bibitem[{F. {van Leeuwen}(2007){van Leeuwen}}]{vanLeeuwen2007}
{van Leeuwen}, F. 2007, \bibinfo{title}{{Validation of the new Hipparcos
  reduction},} \aap, 474, 653, \dodoi{10.1051/0004-6361:20078357}

\bibitem[{F. {Verbunt} \& C. {Zwaan}(1981){Verbunt} \&
  {Zwaan}}]{Verbunt1981A&A}
{Verbunt}, F., \& {Zwaan}, C. 1981, \bibinfo{title}{{Magnetic braking in
  low-mass X-ray binaries.},} \aap, 100, L7

\bibitem[{P. Virtanen {et~al.}(2020)Virtanen, Gommers, Oliphant, Haberland,
  Reddy, Cournapeau, Burovski, Peterson, Weckesser, Bright, {van der Walt},
  Brett, Wilson, Millman, Mayorov, Nelson, Jones, Kern, Larson, Carey, Polat,
  Feng, Moore, {VanderPlas}, Laxalde, Perktold, Cimrman, Henriksen, Quintero,
  Harris, Archibald, Ribeiro, Pedregosa, {van Mulbregt}, \& {SciPy 1.0
  Contributors}}]{2020SciPy-NMeth}
Virtanen, P., Gommers, R., Oliphant, T.~E., {et~al.} 2020,
  \bibinfo{title}{{{SciPy} 1.0: Fundamental Algorithms for Scientific Computing
  in Python},} Nature Methods, 17, 261, \dodoi{10.1038/s41592-019-0686-2}

\bibitem[{W.~D. {Vousden} {et~al.}(2016){Vousden}, {Farr}, \&
  {Mandel}}]{Vousden2016}
{Vousden}, W.~D., {Farr}, W.~M., \& {Mandel}, I. 2016, \bibinfo{title}{{Dynamic
  temperature selection for parallel tempering in Markov chain Monte Carlo
  simulations},} \mnras, 455, 1919, \dodoi{10.1093/mnras/stv2422}

\bibitem[{S. {Wang} {et~al.}(2024){Wang}, {Zhao}, {Feng}, {Ge}, {Shao}, {Cui},
  {Gao}, {Zhang}, {Wang}, {Li}, {Bai}, {Yuan}, {Huang}, {Yuan}, {Zhang}, {Yi},
  {Xiang}, {Li}, {Li}, {Zhang}, {Zhang}, {Han}, {Fan}, {Li}, {Chen}, {Liu},
  {Meng}, {Liu}, {Zhang}, {Gu}, \& {Liu}}]{Wang2024NatAs_G2546}
{Wang}, S., {Zhao}, X., {Feng}, F., {et~al.} 2024, \bibinfo{title}{{A potential
  mass-gap black hole in a wide binary with a circular orbit},} Nature
  Astronomy, 8, 1583, \dodoi{10.1038/s41550-024-02359-9}

\bibitem[{G.-Y. {Xiao} {et~al.}(2023){Xiao}, {Liu}, {Teng}, {Wang}, {Brandt},
  {Zhao}, {Zhao}, {Zhai}, \& {Gao}}]{Xiao2023}
{Xiao}, G.-Y., {Liu}, Y.-J., {Teng}, H.-Y., {et~al.} 2023, \bibinfo{title}{{The
  Masses of a Sample of Radial-velocity Exoplanets with Astrometric
  Measurements},} Research in Astronomy and Astrophysics, 23, 055022,
  \dodoi{10.1088/1674-4527/accb7e}

\bibitem[{G.-Y. {Xiao} {et~al.}(2024){Xiao}, {Feng}, {Shectman}, {Tinney},
  {Teske}, {Carter}, {Jones}, {Wittenmyer}, {D{\'\i}az}, {Crane}, {Wang},
  {Bailey}, {O'Toole}, {Feinstein}, {Rice}, {Essack}, {Montet}, {Shporer}, \&
  {Butler}}]{Xiao2024MNRAS}
{Xiao}, G.-Y., {Feng}, F., {Shectman}, S.~A., {et~al.} 2024,
  \bibinfo{title}{{HD 222237 b: a long-period super-Jupiter around a nearby
  star revealed by radial-velocity and Hipparcos-Gaia astrometry},} \mnras,
  534, 2858, \dodoi{10.1093/mnras/stae2151}

\bibitem[{H.-S. {Xu} {et~al.}(2025){Xu}, {Han}, {Na}, {Wang}, {Ma}, \&
  {Zhu}}]{Xu_CYAri_2025ApJ}
{Xu}, H.-S., {Han}, Z.-T., {Na}, W.-w., {et~al.} 2025, \bibinfo{title}{{The
  Crucial Role of Third Body in Formation and Evolution of Contact Binary
  Systems: Evidence from CY Ari and IK Lyn},} \apj, 988, 85,
  \dodoi{10.3847/1538-4357/ade152}

\bibitem[{D.~G. {York} {et~al.}(1976){York}, {Flannery}, \&
  {Bahcall}}]{York1976ApJ}
{York}, D.~G., {Flannery}, B., \& {Bahcall}, J. 1976,
  \bibinfo{title}{{Circumstellar matter in the binary V Puppis.},} \apj, 210,
  143, \dodoi{10.1086/154812}

\bibitem[{M. {Zechmeister} \& M. {K{\"u}rster}(2009){Zechmeister} \&
  {K{\"u}rster}}]{Zechmeister2009}
{Zechmeister}, M., \& {K{\"u}rster}, M. 2009, \bibinfo{title}{{The generalised
  Lomb-Scargle periodogram. A new formalism for the floating-mean and Keplerian
  periodograms},} \aap, 496, 577, \dodoi{10.1051/0004-6361:200811296}

\end{thebibliography}
\bibliographystyle{aasjournalv7}



\end{document}